# General Relativity, Maxwell's Electrodynamics, and the Foundations of the Quantum Theory of Gravitation and Matter

**Jay R. Yablon**[*]
910 Northumberland Drive
Schenectady, New York, 12309-2814

**Abstract:**

**The formalism of electric – magnetic duality, first pioneered by Reinich and Wheeler, extends General Relativity to encompass non-Abelian fields. Several energy tensors $T^{\mu\nu}$ with non-vanishing trace matter are developed as a function solely of the field strength tensor $F^{\mu\nu}$, including the Euler tensor, and tensors for matter in flux, pressure in flux, and stationary pressure. The spacetime metric $g_{\mu\nu}$ is not only a solution to the second-order Einstein equation based on $T^{\mu\nu}$, but is also constrained by a third-order equation involving the Bianchi identity together with the gravitational energy components $\kappa_\nu$ for each $T^{\mu\nu}$. The common appearance of $F^{\mu\nu}$ in all of the $T^{\mu\nu}$ and $\kappa_\nu$ makes it possible to obtain quantum solutions for the spacetime metric, thereby geometrizing quantum physics as a non-linear theory.**

---

[*] jyablon@nycap.rr.com

# 1. Introduction

In his 1916 paper on *The Foundation of the General Theory of Relativity* [1], Albert Einstein demonstrates the conservation of energy by relating the total energy tensor $T^\mu{}_\nu$ to the Bianchi identity $\left(R^\mu{}_\nu - \tfrac{1}{2}\delta^\mu{}_\nu R\right)_{;\mu} = 0$, the Maxwell energy tensor $T^\mu{}_{\nu\,Maxwell}$, the field strength tensor $F^{\mu\nu}$, and the energy tensor $t^\mu{}_\nu$ of the gravitational field according to:

$$
\begin{aligned}
-\kappa T^\mu{}_{\nu;\mu} &= -\kappa\left(T^\mu{}_{\nu\,Maxwell} + t^\mu{}_\nu\right)_{;\mu} = \left(R^\mu{}_\nu - \tfrac{1}{2}\delta^\mu{}_\nu R\right)_{;\mu} = 0 \\
&= \tfrac{\kappa}{4\pi}\left[\tfrac{1}{2}F^{u\sigma}\left(F_{\mu\nu;o} + F_{\nu\sigma;\mu} + F_{\sigma\mu;\nu}\right)\right] = \tfrac{\kappa}{4\pi}\left[\left(F^{u\sigma}F_{\nu\sigma} - \tfrac{1}{4}\delta^\mu{}_\nu F^{\tau\sigma}F_{\tau\sigma}\right)_{;\mu} - F_{\nu\mu}F^{\sigma\mu}{}_{;\sigma}\right]. \\
&= \tfrac{\kappa}{4\pi}\left[\tfrac{1}{2}\left(F^{u\sigma}F_{\nu\sigma} + {*F^{u\sigma}}{*F_{\nu\sigma}}\right)_{;\mu} - F_{\nu\mu}F^{\sigma\mu}{}_{;\sigma}\right] = 0
\end{aligned}
\qquad(1.1)
$$

The "dual" of the field strength tensor above is defined as $*F^{\sigma\tau} \equiv \tfrac{1}{2!}\varepsilon^{\delta\gamma\sigma\tau}F_{\delta\gamma}$ using the Levi-Civita formalism, see, for example, [2], [3], [4] at pages 87-89, and [5], equations (5.4) and (5.6). This also employs $\varepsilon^{\delta\mu\sigma}\varepsilon_{\alpha\beta\nu\sigma} = -\delta^{\delta\mu\sigma}{}_{\alpha\beta\nu\sigma}$, see [4] at pages 87-89, and [5] following (3.1). Integral to the identity of $T^\mu{}_{\nu;\mu}$ with zero and thus to energy conservation is the second of Maxwell's equations:

$$
\tfrac{1}{4\pi}\left(F_{\tau\sigma;\nu} + F_{\sigma\nu;\tau} + F_{\nu\tau;\sigma}\right) = 0,
\qquad(1.2)
$$

which in turn has its identity to zero ensured by the Abelian relationship:

$$
F_{\mu\nu} = A_{\nu;\mu} - A_{\mu;\nu}
\qquad(1.3)
$$

between the four-vector potential $A^u$ and $F^{uv}$. Absent (1.3) above, or, if (1.3) above were to instead be replaced by the non-Abelian (Yang-Mills) relationship of the general form:

$$
F_{i\mu\nu} = A_{i\nu;\mu} - A_{i\mu;\nu} - gf_{ijk}A^j{}_\mu A^k{}_\nu,
\qquad(1.4)
$$

where i is an internal symmetry index, $f_{ijk}$ are group structure constants, and g is an interaction charge, then (1.2) would no longer be assured to vanish identically, and so the total energy tensor as specified in (1.1) would no longer be assured to be conserved, $T^\mu{}_{\nu;\mu} \neq 0$. More to the point, the total energy $T^\mu{}_\nu$ would no longer be "total," but would need to be exchanged with additional energy terms not appearing in (1.1). It is to be observed that non-linear $\mathbf{A}\cdot\mathbf{A}$ interaction terms such as in (1.4) are also central to modern particle physics, and so must eventually be accommodated by an equation of the form (1.1) if we are ever to understand weak and strong quantum interactions in a gravitational, geometrodynamic framework.

The set of connections in (1.1) do, of course, underlie the successful identification of the Maxwell – Poynting tensor for "matter" with the integrable terms in (1.1), according to:

$$
T^\mu{}_{\nu\,Maxwell} \equiv -\tfrac{1}{4\pi}\left[F^{u\sigma}F_{\nu\sigma} - \tfrac{1}{4}\delta^\mu{}_\nu F^{\tau\sigma}F_{\tau\sigma}\right] = -\tfrac{1}{8\pi}\left[F^{u\sigma}F_{\nu\sigma} + {*F^{u\sigma}}{*F_{\nu\sigma}}\right],
\qquad(1.5)
$$

as well as the identification of the non-integrable energy tensor $t^\mu{}_\nu$ of the "gravitational field":

$$\kappa_\nu \equiv t^\mu{}_{\nu;\mu} = \tfrac{1}{4\pi} F_{\mu\nu} F^{\sigma\mu}{}_{;\sigma} = F_{\mu\nu} J^\mu, \tag{1.6}$$

which represents the density of energy-momentum exchanged per unit of time, between the electric current density $J^\mu$ and electromagnetic field $F_{\mu\nu}$ (see [1], following equation (65a)). In the above, we have employed Maxwell's remaining equation:

$$J^\nu = \tfrac{1}{4\pi} F^{\mu\nu}{}_{;\mu}. \tag{1.7}$$

However, if we set:

$$-\kappa T^\mu{}_{\nu\,Maxwell} = R^\mu{}_\nu - \tfrac{1}{2}\delta^\mu{}_\nu R = \tfrac{\kappa}{4\pi}\left(F^{u\sigma} F_{\nu\sigma} - \tfrac{1}{4}\delta^\mu{}_\nu F^{\tau\sigma} F_{\tau\sigma}\right) = \tfrac{\kappa}{8\pi}\left(F^{u\sigma} F_{\nu\sigma} + {}^*F^{u\sigma}{}^*F_{\nu\sigma}\right), \tag{1.8}$$

then, on account of (1.1), we find that $\kappa_\nu = 0$ in (1.6) and so the current is thought to vanish, $J^\mu = 0$. Additionally, the trace equation vanishes:

$$\kappa T_{Maxwell} = R = -\tfrac{\kappa}{4\pi}\left(F^{\mu\sigma} F_{\mu\sigma} - \tfrac{1}{4}\delta^\mu{}_\mu F^{\tau\sigma} F_{\tau\sigma}\right) = -\tfrac{1}{8\pi}\left(F^{\tau\sigma} F_{\tau\sigma} + {}^*F^{\tau\sigma}{}^*F_{\tau\sigma}\right) = 0, \tag{1.9}$$

on account of the photon mediators of the electromagnetic interaction being massless, and therefore traveling at the speed of light. Thus, as stated by Einstein in 1919, "we cannot arrive at a theory of the electron [and matter generally] by restricting ourselves to the electromagnetic components of the Maxwell-Lorentz theory, as has long been known." ([6] at page 192)

In addition to the problem of matter, there are other problems which arise from equation (1.1). Because (1.1) relies upon the Abelian field (1.3), it is simply not valid for non-Abelian fields. Thus, without a reconsideration of (1.1), one cannot apply the General Theory of Relativity to non-Abelian interactions. This immediately bars understanding $SU(2)_W$ weak interactions, or $SU(3)_{QCD}$ interactions, for example, in connection with Einstein's theory of gravitation.

Additionally, (1.1) excludes, *a priori*, the possibility that magnetic monopoles might actually exist in nature, whether of an ordinary magnetic, or "chromo" magnetic character.[*] In particular, if we define the third-rank antisymmetric tensor:

$$P_{\tau\sigma\nu} \equiv \tfrac{1}{4\pi}\left(F_{\tau\sigma;\nu} + F_{\sigma\nu;\tau} + F_{\nu\tau;\sigma}\right), \tag{1.10}$$

and because the current four-vector for magnetic charges may be specified in terms of $P_{\tau\sigma\nu}$ and ${}^*F^{\mu\nu}$ by (see, for example, [4], at pages 87-89, and [5], equations (5.4) and (5.6)):

$$P^\sigma = {}^*P^\sigma = \tfrac{1}{3!}\varepsilon^{\alpha\tau\gamma\sigma} P_{\alpha\tau\gamma} = \tfrac{1}{4\pi} {}^*F^{\mu\nu}{}_{;\mu}, \tag{1.11}$$

---

[*] We take the term "chromo-magnetic" as applying broadly to the magnetic monopoles of any non-Abelian interaction, not just the strong QCD interactions, and ordinary "magnetic" to apply particularly to $U(1)_{EM}$.

we see that (1.1), as it stands, expressly *forecloses* the existence of magnetic monopoles, because the vanishing of $P_{\tau\sigma\nu}$ in (1.10) causes $P^\sigma$ in (1.11) to vanish as well. * Any theory which allows magnetic (or "chromo" magnetic) monopoles by using a non-Abelian field (1.4), requires that (1.1) be suitably-modified for total energy to be properly conserved, because $(F_{\mu\nu;\sigma} + F_{\nu\sigma;\mu} + F_{\sigma\mu;\nu})$ will no longer be identical to zero. For completeness, we also define (see [5], equations (3.17) and (4.5)):

$$J_{\tau\sigma\nu} \equiv -\tfrac{1}{4\pi}\left(*F_{\tau\sigma;\nu} + *F_{\sigma\nu;\tau} + *F_{\nu\tau;\sigma}\right) = \tfrac{1}{3!}\varepsilon_{\gamma\tau\sigma\nu}J^\gamma. \tag{1.12}$$

As we shall demonstrate, all of these problems stem from the fact that (1.1) relies upon the vanishing of the antisymmetric combination of terms in (1.2) to enforce the conservation of total energy. The term $T^\mu{}_{\nu;\mu} = 0$ is solidly-grounded: it is the quintessential statement that total energy must be conserved. The Bianchi identity $\left(R^\mu{}_\nu - \tfrac{1}{2}\delta^\mu{}_\nu R\right)_{;\mu} = 0$ is equally solid: although one can also add a "cosmological" term $\left(R^\mu{}_\nu - \tfrac{1}{2}\delta^\mu{}_\nu R + \Lambda\delta^\mu{}_\nu\right)_{;\mu} = 0$, one is assured by the very nature of Riemannian geometry that either combination of terms will always be zero. Not so, however, for $\tfrac{1}{2}F^{\mu\sigma}(F_{\mu\nu;\sigma} + F_{\nu\sigma;\mu} + F_{\sigma\mu;\nu}) = 0$. This term relies directly on the Abelian field (1.3) and on the supposition that magnetic monopoles (1.11) vanish. Absent this supposition, $T^\mu{}_\nu$ is no longer conserved, and so can no longer be regarded as the "total" energy tensor.

To remedy this, thereby generalizing the General Theory of Relativity to non-Abelian interactions and permitting the existence of (chromo) magnetic monopoles, we need to find a set of "electrodynamic Bianchi identities" constructed solely out of the field strength tensor $F^{\mu\nu}$ and its dual $*F^{\mu\nu}$, which are analogous to the identity $\tfrac{1}{2}F^{\mu\sigma}(F_{\mu\nu;\sigma} + F_{\nu\sigma;\mu} + F_{\sigma\mu;\nu}) = 0$, but which are *totally independent* of whether $F^{\mu\nu}$ has the Abelian form $F_{\mu\nu} = A_{\nu;\mu} - A_{\mu;\nu}$ of (1.3), the non-Abelian form $F_{i\mu\nu} = A_{i\nu;\mu} - A_{i\mu;\nu} - gf_{ijk}A^j{}_\mu A^k{}_\nu$ of (1.4), or, indeed, any other form, so long as it remains an antisymmetric tensor.

## 2. The Energy Tensor of Trace Matter

The "electrodynamic Bianchi identities" referred to above will be based on the identity:

$$\tfrac{1}{2}A^{\sigma\tau}\left(B_{\tau\sigma;\nu} + B_{\sigma\nu;\tau} + B_{\nu\tau;\sigma}\right) - *A_{\nu\sigma}*B^{\tau\sigma}{}_{;\tau} = 0. \tag{2.1}$$

This identity holds for *any* two antisymmetric tensors $*A_{\mu\nu} \equiv \tfrac{1}{2!}\varepsilon_{\sigma\tau\mu\nu}A^{\sigma\tau}$ and $*B_{\mu\nu} \equiv \tfrac{1}{2!}\varepsilon_{\sigma\tau\mu\nu}B^{\sigma\tau}$ in $\Re^4$, see, e.g., [3] at page 251 and the discussion following (2.1) in [5]. It is a useful exercise to confirm this identity using $\varepsilon^{\delta\eta\mu\sigma}\varepsilon_{\alpha\beta\nu\sigma} = -\delta^{\delta\eta\mu\sigma}{}_{\alpha\beta\nu\sigma}$, and observe that this identity is based solely on the antisymmetric character of $A^{\sigma\tau}$ and $B^{\tau\sigma}$ and the definitions of their duals. What is especially important is that this identity is thereby *independent* of whether $A^{\sigma\tau}$ and $B^{\tau\sigma}$ are tensors of an Abelian or non-Abelian field theory. In the latter case, this is easily generalized to $\tfrac{1}{2}A^{i\sigma\tau}(B_{i\tau\sigma;\nu} + B_{i\sigma\nu;\tau} + B_{i\nu\tau;\sigma}) - *A_{i\nu\sigma}*B^{i\tau\sigma}{}_{;\tau} = 0$, where i is the internal symmetry index of the particular non-Abelian, Yang Mills gauge groups under consideration, such as $SU(2)_W$ or $SU(3)_{QCD}$. For the moment, we shall consider the Abelian interaction, $U(1)_{EM}$.

---

* The author explores this in depth in [5].

If we write identity (2.1) specifically in terms of the field strength tensor $F^{\sigma\tau}$, that is, if we set $A^{\sigma\tau} = B^{\sigma\tau} = F^{\sigma\tau}$, then (2.1) becomes (see, e.g., [3] at page 251, note 22):

$$\tfrac{1}{2} F^{\sigma\tau}\left(F_{\tau\sigma;\nu} + F_{\sigma\nu;\tau} + F_{\nu\tau;\sigma}\right) - {}^*F_{\nu\sigma}{}^*F^{\tau\sigma}{}_{;\tau} = 0. \tag{2.2}$$

If apply a discrete global duality transformation $F^{\sigma\tau} \to {}^*F^{\sigma\tau}$ to the $F^{\sigma\tau}$ in (2.2), then because ${}^{**}=-1$, we can derive the related (not independent), duality-rotated identity:

$$\tfrac{1}{2} {}^*F^{\sigma\tau}\left({}^*F_{\tau\sigma;\nu} + {}^*F_{\sigma\nu;\tau} + {}^*F_{\nu\tau;\sigma}\right) - F_{\nu\sigma} F^{\tau\sigma}{}_{;\tau} = 0. \tag{2.3}$$

Because $A^{\sigma\tau}$ and $B^{\sigma\tau}$ in (2.1) are *independent* tensors, again using ${}^{**}=-1$, we may also set $A^{\sigma\tau} = F^{\sigma\tau}$ and $B^{\sigma\tau} = {}^*F^{\sigma\tau}$, to yield:

$$\tfrac{1}{2} F^{\sigma\tau}\left({}^*F_{\tau\sigma;\nu} + {}^*F_{\sigma\nu;\tau} + {}^*F_{\nu\tau;\sigma}\right) + {}^*F_{\nu\sigma} F^{\tau\sigma}{}_{;\tau} = 0. \tag{2.4}$$

This identity cannot be arrived at from either of (2.2) or (2.3) by a duality transformation, and so is *independent* of (2.2) and (2.3). Finally, if we apply $F^{\sigma\tau} \to {}^*F^{\sigma\tau}$ uniformly to $F^{\sigma\tau}$ in the above, we also derive the related, not independent:

$$-\tfrac{1}{2} {}^*F^{\sigma\tau}\left(F_{\tau\sigma;\nu} + F_{\sigma\nu;\tau} + F_{\nu\tau;\sigma}\right) - F_{\nu\sigma} {}^*F^{\tau\sigma}{}_{;\tau} = 0. \tag{2.5}$$

where we have left in (2.5), the minus signs that result from applying $F^{\sigma\tau} \to {}^*F^{\sigma\tau}$ to (2.4).

Any one of the identities (2.2), (2.3), (2.4) and (2.5) has the very-desirable property that it is identically equal to zero, again, whether the field strength tensor is Abelian, $F_{\mu\nu}$ as in (1.3), or non-Abelian $F_{i\mu\nu}$ as in (1.4). This means also, that these identities do not depend upon equation (1.2), $\tfrac{1}{4\pi}\left(F_{\tau\sigma;\nu} + F_{\sigma\nu;\tau} + F_{\nu\tau;\sigma}\right) = 0$, and so in general can accommodate non-vanishing magnetic monopole charges (1.10), (1.11), which are thus permitted, but not required. These identities have the further virtue that they are constructed solely from the field strength tensor $F^{\mu\nu}$ and its dual ${}^*F^{\mu\nu}$, and, comparing to (1.1), that they are of the correct rank and differential order to be related to the contracted divergence of the energy tensor and the Bianchi identity, i.e., to $-\kappa T^\mu{}_{\nu;\mu} = \left(R^\mu{}_\nu - \tfrac{1}{2}\delta^\mu{}_\nu R\right)_{;\mu} = 0$. Now, we return to consider equation (1.1) in light of these identities.

Identity (2.2)

Let us first consider identity (2.2), because this includes the problematic term $\tfrac{1}{2} F^{\mu\sigma}\left(F_{\mu\nu;\sigma} + F_{\nu\sigma;\mu} + F_{\sigma\mu;\nu}\right)$ which was used to conserve energy in (1.1). First, let us write (2.2) in several alternative forms (deriving these alternative formulations is a good exercise):

$$\frac{1}{2} F^{\mu\sigma} \left( F_{\mu\nu;\sigma} + F_{\nu\sigma;\mu} + F_{\sigma\mu;\nu} \right) - {}^*F_{\nu\mu} {}^* F^{\sigma\mu}{}_{;\sigma}$$
$$= \left( F^{u\sigma} F_{\nu\sigma} - \tfrac{1}{4} \delta^{\mu}{}_{\nu} F^{\tau\sigma} F_{\tau\sigma} \right)_{;\mu} - F_{\nu\mu} F^{\sigma\mu}{}_{;\sigma} - {}^*F_{\nu\mu} {}^* F^{\sigma\mu}{}_{;\sigma} \qquad (2.6)$$
$$= \tfrac{1}{2} \left( F^{u\sigma} F_{\nu\sigma} + {}^*F^{u\sigma} {}^* F_{\nu\sigma} \right)_{;\mu} - F_{\nu\mu} F^{\sigma\mu}{}_{;\sigma} - {}^*F_{\nu\mu} {}^* F^{\sigma\mu}{}_{;\sigma} = 0$$

Because this is a mathematical identity based solely on the definition ${}^*F^{\sigma\tau} \equiv \frac{1}{2!} \varepsilon^{\delta\gamma\sigma\tau} F_{\delta\gamma}$ as well as the antisymmetric character of $F^{\mu\nu}$, this identity is equal to zero for Abelian and non-Abelian field theory alike, and thus, does not exclude magnetic monopoles. In fact, a magnetic monopole current does appear explicitly in (2.6) via the term ${}^*F_{\mu\nu} {}^* F^{\sigma\mu}{}_{;\sigma} = 4\pi {}^* F_{\mu\nu} P^\mu$, see (1.11). One may therefore use $\frac{1}{2} F^{\mu\sigma} \left( F_{\mu\nu;\sigma} + F_{\nu\sigma;\mu} + F_{\sigma\mu;\nu} \right) - {}^*F_{\nu\mu} {}^* F^{\sigma\mu}{}_{;\sigma} = 0$ in place of the Abelian-dependent $\frac{1}{2} F^{u\sigma} \left( F_{\mu\nu;o} + F_{\nu\sigma;\mu} + F_{\sigma\mu;\nu} \right) = 0$ in (1.1) to write:

$$\begin{aligned}
-\kappa T^\mu{}_{\nu;\mu} &= -\kappa \left( T^\mu{}_{\nu\,Maxwell} + t^\mu{}_\nu \right)_{;\mu} = \left( R^\mu{}_\nu - \tfrac{1}{2} \delta^\mu{}_\nu R \right)_{;\mu} = 0 \\
&= \tfrac{\kappa}{4\pi} \left[ \tfrac{1}{2} F^{\mu\sigma} \left( F_{\mu\nu;\sigma} + F_{\nu\sigma;\mu} + F_{\sigma\mu;\nu} \right) - {}^*F_{\nu\mu} {}^* F^{\sigma\mu}{}_{;\sigma} \right] \\
&= \tfrac{\kappa}{4\pi} \left[ \tfrac{1}{2} \left( F^{u\sigma} F_{\nu\sigma} + {}^*F^{u\sigma} {}^* F_{\nu\sigma} \right)_{;\mu} - F_{\nu\mu} F^{\sigma\mu}{}_{;\sigma} - {}^*F_{\nu\mu} {}^* F^{\sigma\mu}{}_{;\sigma} \right] \\
&= \tfrac{\kappa}{4\pi} \left[ \left( F^{u\sigma} F_{\nu\sigma} - \tfrac{1}{4} \delta^\mu{}_\nu F^{\tau\sigma} F_{\tau\sigma} \right)_{;\mu} - F_{\nu\mu} F^{\sigma\mu}{}_{;\sigma} - {}^*F_{\nu\mu} {}^* F^{\sigma\mu}{}_{;\sigma} \right] = 0
\end{aligned} \qquad (2.7)$$

In the above, energy is assured to be conserved *irrespective of the Abelian or non-Abelian nature of $F^{\mu\nu}$*. Even if $\left( F_{\mu\nu;\sigma} + F_{\nu\sigma;\mu} + F_{\sigma\mu;\nu} \right) \neq 0$ in the above, we still have $T^\mu{}_{\nu;\mu} = 0$. For non-Abelian fields, the core identity is $\frac{1}{2} F^{i\mu\sigma} \left( F_{i\mu\nu;\sigma} + F_{i\nu\sigma;\mu} + F_{i\sigma\mu;\nu} \right) - {}^*F_{i\nu\mu} {}^* F^{i\sigma\mu}{}_{;\sigma} = 0$. The Maxwell tensor $T^\mu{}_{\nu\,Maxwell} = -\frac{1}{4\pi} \left[ F^{u\sigma} F_{\nu\sigma} - \tfrac{1}{4} \delta^\mu{}_\nu F^{\tau\sigma} F_{\tau\sigma} \right]$ of (1.5) for the "matter" of the electromagnetic field emerges unchanged, in the integrable portion of the above. However, the exchange of energy between the energy tensor $t^\mu{}_\nu$ of the gravitational field and the electromagnetic field, earlier shown in (1.6), now takes on the slightly altered form:

$$\kappa_\nu = t^\mu{}_{\nu;\mu} = \tfrac{1}{4\pi} \left( F_{\nu\mu} F^{\sigma\mu}{}_{;\sigma} + {}^*F_{\nu\mu} {}^* F^{\sigma\mu}{}_{;\sigma} \right) = F_{\nu\mu} J^\mu + {}^*F_{\nu\mu} P^\mu . \qquad (2.8)$$

This now includes a new term ${}^*F_{\nu\mu} P^\mu$ which contains a magnetic monopole current $P^\mu$. As soon as we generalize to accommodate non-Abelian fields and allow for the non-vanishing of magnetic monopoles, the first thing we find out is that these magnetic monopoles, if they exist, contribute to the gravitational field energy right alongside electric charges.

However, the problem still remains, that if we set $-\kappa T^\mu{}_{\nu\,Maxwell} = R^\mu{}_\nu - \tfrac{1}{2} \delta^\mu{}_\nu R$ in (2.7), as in (1.8), we still end up with $\kappa_\nu = F_{\nu\mu} J^\mu + {}^*F_{\nu\mu} P^\mu = 0$, as well as the zero trace $\kappa T = R = 0$ of (1.9). While $J^\mu$ does not necessarily vanish, it will seem to vanish wherever the magnetic monopoles vanish, that is, wherever $P^\mu = 0$. So, we still have the problem of finding a suitable energy tensor for non-vanishing trace matter. Let us see if one of the other identities (2.3) through (2.5) might be of help.

## Identity (2.3)

We next turn to (2.3), which is (2.2) following the discrete global duality transformation $F^{\sigma\tau} \to *F^{\sigma\tau}$. The $\frac{1}{2}*F^{\sigma\tau}\left(*F_{\tau\sigma;\nu}+*F_{\sigma\nu;\tau}+*F_{\nu\tau;\sigma}\right)$ can be written in the alternative formulations:

$$\frac{1}{2}*F^{\mu\sigma}\left(*F_{\mu\nu;\sigma}+*F_{\nu\sigma;\mu}+*F_{\sigma\mu;\nu}\right)=\left(*F^{\mu\sigma}*F_{\nu\sigma}-\tfrac{1}{4}\delta^{\mu}{}_{\nu}*F^{\sigma\tau}*F_{\sigma\tau}\right)_{;\mu}-*F_{\nu\mu}*F^{\sigma\mu}{}_{;\sigma}$$
$$=\left(-\tfrac{1}{4}\delta^{\mu}{}_{\nu}F^{\sigma\tau}F_{\sigma\tau}-\tfrac{1}{4}\delta^{\mu}{}_{\nu}*F^{\sigma\tau}*F_{\sigma\tau}\right)_{;\mu}-*F_{\nu\mu}*F^{\sigma\mu}{}_{;\sigma}=-*F_{\nu\mu}*F^{\sigma\mu}{}_{;\sigma} \quad (2.9)$$

The term $*F^{\mu\sigma}*F_{\nu\sigma}-\tfrac{1}{4}\delta^{\mu}{}_{\nu}*F^{\sigma\tau}*F_{\sigma\tau}$ resembles the term $F^{\mu\sigma}F_{\nu\sigma}-\tfrac{1}{4}\delta^{\mu}{}_{\nu}F^{\tau\sigma}F_{\tau\sigma}$ which is at the heart of the Maxwell tensor. But, when the dual term $*F^{\mu\sigma}*F_{\nu\sigma}$ is rewritten in terms of $F^{\sigma\tau}$,[*] then on account of the vanishing trace relationship $F^{\tau\sigma}F_{\tau\sigma}+*F^{\tau\sigma}*F_{\tau\sigma}=0$, (1.9), the term $*F^{\mu\sigma}*F_{\nu\sigma}-\tfrac{1}{4}\delta^{\mu}{}_{\nu}*F^{\sigma\tau}*F_{\sigma\tau}=0$, identically. Thus, identity (2.3) can be written simply as:

$$\tfrac{1}{2}*F^{\mu\sigma}\left(*F_{\mu\nu;\sigma}+*F_{\nu\sigma;\mu}+*F_{\sigma\mu;\nu}\right)-F_{\nu\mu}F^{\sigma\mu}{}_{;\sigma}=-F_{\nu\mu}F^{\sigma\mu}{}_{;\sigma}+*F_{\nu\mu}*F^{\sigma\mu}{}_{;\sigma}=0. \quad (2.10)$$

If we were, therefore, to use this identity to ensure conservation, with a different energy tensor which we shall write $T^{\mu}{}_{\nu Vacuum}$ we would write (2.7) as:

$$-\kappa T^{\mu}{}_{\nu;\mu}=-\kappa\left(T^{\mu}{}_{\nu Vacuum}+t^{\mu}{}_{\nu}\right)_{;\mu}=\left(R^{\mu}{}_{\nu}-\tfrac{1}{2}\delta^{\mu}{}_{\nu}R\right)_{;\mu}=0$$
$$=\tfrac{\kappa}{4\pi}\left[\tfrac{1}{2}*F^{\mu\sigma}\left(*F_{\mu\nu;\sigma}+*F_{\nu\sigma;\mu}+*F_{\sigma\mu;\nu}\right)-F_{\nu\mu}F^{\sigma\mu}{}_{;\sigma}\right] \quad (2.11)$$
$$=\tfrac{\kappa}{4\pi}\left[F_{\mu\nu}F^{\sigma\mu}{}_{;\sigma}+*F_{\mu\nu}*F^{\sigma\mu}{}_{;\sigma}\right]=0$$

and would therefore find that under the transformation from $F^{\sigma\tau}\to*F^{\sigma\tau}$, the resulting energy tensor vanishes, $T^{\mu}{}_{\nu Vacuum}=0$, which makes plain why we have labeled that energy tensor as such. Despite the vanishing of $T^{\mu}{}_{\nu Vacuum}$, the gravitational components of (2.8) emerge intact, but as $\kappa_{\nu}=t^{\mu}{}_{\nu;\mu}=0$. Apparently, the Maxwell tensor (1.5) and the tensor of the vacuum, $T^{\mu}{}_{\nu Vacuum}=0$, have the same gravitational components $\kappa_{\nu}$. The equation $\kappa_{\nu}=t^{\mu}{}_{\nu;\mu}=0$ expresses the separate conservation of "gravitational field" energy in a vacuum devoid of "matter."[**] The $J^{\mu}$ in the absence of $P^{\mu}$ still appears to vanish, and $T^{\mu}{}_{\nu Vacuum}=0$ of course is traceless, like the Maxwell tensor. The problem of matter still persists.

## Identity (2.5)

It will be convenient if we next explore the *independent* identity (2.5), and then return to (2.4). Using $\left(*F^{\mu\sigma}F_{\nu\sigma}\right)_{;\mu}=*F^{\mu\sigma}{}_{;\mu}F_{\nu\sigma}+*F^{\mu\sigma}F_{\nu\sigma;\mu}$, the term $-\tfrac{1}{2}*F^{\sigma\tau}\left(F_{\tau\sigma;\nu}+F_{\sigma\nu;\tau}+F_{\nu\tau;\sigma}\right)$ in (2.5) has the alternative formulation:

---

[*] $*F^{\mu\sigma}*F_{\nu\sigma}=\tfrac{1}{4}\varepsilon^{\alpha\beta\mu\sigma}F_{\alpha\beta}\varepsilon_{\delta\gamma\nu\sigma}F^{\delta\gamma}=-\tfrac{1}{4}\delta^{\alpha\beta\mu\sigma}_{\delta\gamma\nu\sigma}F_{\alpha\beta}F^{\delta\gamma}=-\tfrac{1}{4}\delta^{\mu}{}_{\nu}F_{\delta\gamma}F^{\delta\gamma}$

[**] See [1], start of §14, where the distinction between "gravitational field" and "matter" is discussed.

$$-\tfrac{1}{2} * F^{\mu\sigma}\left(F_{\mu\nu;\sigma} + F_{\nu\sigma;\mu} + F_{\sigma\mu;\nu}\right)$$
$$= -\left(* F^{\mu\sigma} F_{\nu\sigma} - \tfrac{1}{2}\delta^{\mu}{}_{\nu} * F^{\sigma\tau} F_{\sigma\tau}\right)_{;\mu} + F_{\nu\sigma} * F^{\mu\sigma}{}_{;\mu} - \tfrac{1}{2} F^{\sigma\tau} * F_{\sigma\tau;\nu} \quad (2.12)$$

and so identity (2.5) may be alternatively expressed as:

$$-\tfrac{1}{2} * F^{\mu\sigma}\left(F_{\mu\nu;\sigma} + F_{\nu\sigma;\mu} + F_{\sigma\mu;\nu}\right) - F_{\nu\sigma} * F^{\mu\sigma}{}_{;\mu}$$
$$= \left(* F^{\mu\sigma} F_{\nu\sigma} - \tfrac{1}{2}\delta^{\mu}{}_{\nu} * F^{\sigma\tau} F_{\sigma\tau}\right)_{;\mu} - \tfrac{1}{2} F^{\sigma\tau} * F_{\sigma\tau;\nu} = 0 \quad (2.13)$$

It is very significant that in the course of separating integrable from non-integrable terms, we end up with a $\tfrac{1}{2}\delta^{\mu}{}_{\nu}$ rather than a $\tfrac{1}{4}\delta^{\mu}{}_{\nu}$. If we were, therefore, to ensure energy conservation with this identity, we would write (2.7) as:

$$-\kappa T^{\mu}{}_{\nu}(A)_{;\mu} = -\kappa\left(\Im^{\mu}{}_{\nu}(A) + t^{\mu}{}_{\nu}(A)\right)_{;\mu} = \left(R^{\mu}{}_{\nu} - \tfrac{1}{2}\delta^{\mu}{}_{\nu} R\right)_{;\mu} = 0$$
$$= -\tfrac{\kappa}{4\pi}\left[-\tfrac{1}{2} * F^{\mu\sigma}\left(F_{\mu\nu;\sigma} + F_{\nu\sigma;\mu} + F_{\sigma\mu;\nu}\right) - F_{\nu\mu} * F^{\sigma\mu}{}_{;\sigma}\right] \quad (2.14)$$
$$= -\tfrac{\kappa}{4\pi}\left[-\left(* F^{\mu\sigma} F_{\nu\sigma} - \tfrac{1}{2}\delta^{\mu}{}_{\nu} * F^{\sigma\tau} F_{\sigma\tau}\right)_{;\mu} - \tfrac{1}{2} F^{\sigma\tau} * F_{\sigma\tau;\nu}\right] = 0$$

where "A" is a label used to distinguish the above from analogous tensors $T^{\mu}{}_{\nu}(B)$, $\Im^{\mu}{}_{\nu}(B)$, $t^{\mu}{}_{\nu}(B)$ and related objects which will emerge from considering the final identity (2.4). Here, the associations are quite different than in (2.7). The integrable term $\left(* F^{\mu\sigma} F_{\nu\sigma} - \tfrac{1}{2}\delta^{\mu}{}_{\nu} * F^{\sigma\tau} F_{\sigma\tau}\right)$ in the above does not at all resemble the Maxwell tensor or $T^{\mu}{}_{\nu Vacuum} = 0$, and now becomes associated with:

$$\Im^{\mu}{}_{\nu}(A) = -\tfrac{1}{4\pi}\left(* F^{\mu\sigma} F_{\nu\sigma} - \tfrac{1}{2}\delta^{\mu}{}_{\nu} * F^{\sigma\tau} F_{\sigma\tau}\right). \quad (2.15)$$

Similarly, here the gravitational field exchanges energy with matter tensor $\Im^{\mu}{}_{\nu}(A)$ according to:

$$K_{\nu}(A) \equiv t^{\mu}{}_{\nu}(A)_{;\mu} = -\tfrac{1}{8\pi} F^{\sigma\tau} * F_{\sigma\tau;\nu}. \quad (2.16)$$

Equations (2.14) through (2.16) have a number of properties of compelling interest. First, if we regard (2.15) as the source of matter for the Einstein equation by setting (similarly to (1.8), see also [6] at equation (3)):

$$-\kappa T^{\mu}{}_{\nu}(A) = -\kappa \Im^{\mu}{}_{\nu}(A) = \left(R^{\mu}{}_{\nu} - \tfrac{1}{2}\delta^{\mu}{}_{\nu} R\right) = \tfrac{\kappa}{4\pi}\left(* F^{\mu\sigma} F_{\nu\sigma} - \tfrac{1}{2}\delta^{\mu}{}_{\nu} * F^{\sigma\tau} F_{\sigma\tau}\right), \quad (2.17)$$

we find from (2.14) that:

$$K_{\nu}(A) \equiv t^{\mu}{}_{\nu}(A)_{;\mu} = -\tfrac{1}{8\pi} F^{\sigma\tau} * F_{\sigma\tau;\nu} = 0. \quad (2.18)$$

Importantly, when we set this term $\kappa_\nu(A)$ to zero, neither the electric current $J^\mu$ nor the magnetic current $P^\mu$ appears to vanish. Nor does the combination $\kappa_\nu = F_{\nu\mu}J^\mu + {}^*F_{\nu\mu}P^\mu$ of (2.8). So it is possible that $\mathfrak{I}^\mu{}_\nu(A)$ might help us understand the gravitational composition of matter.

Second, equation (2.17), because of its $\frac{1}{2}\delta^\mu{}_\nu$ factor, permits a direct connection between of the Ricci curvature tensor and the electromagnetic field tensors, as such:

$$R_{\mu\nu} = \tfrac{\kappa}{4\pi} {}^*F_{\mu\sigma}F_\nu{}^\sigma = \tfrac{\kappa}{4\pi} g_{\tau\nu} F^{\tau\sigma} {}^*F_{\mu\sigma}, \tag{2.19}$$

and so is suggestive of a very direct connection between electromagnetic fields and the curvature of spacetime. The question that arises from close inspection of the above, is whether the term $\tfrac{\kappa}{4\pi} g_{\tau\nu} F^{\tau\sigma} {}^*F_{\mu\sigma}$ is symmetric under the interchange of the indexes $\mu \Leftrightarrow \nu$, as it must be to have a proper association with $R_{\mu\nu} = R_{\nu\mu}$. As a general rule, if $F^{\tau\sigma}$ and ${}^*F_{\mu\sigma}$ were fully independent tensors, then there would clearly be no reason to expect this term to be symmetric. However, because these are related by ${}^*F^{\sigma\tau} \equiv \tfrac{1}{2!}\varepsilon^{\delta\gamma\sigma\tau}F_{\delta\gamma}$, the possibility exists that this is symmetric, i.e., that $\tfrac{\kappa}{4\pi} g_{\tau\nu} F^{\tau\sigma} {}^*F_{\mu\sigma} = \tfrac{\kappa}{4\pi} g_{\tau\mu} F^{\tau\sigma} {}^*F_{\nu\sigma}$.

It turns out that the above is indeed symmetric under $\mu \Leftrightarrow \nu$ interchange, which can be seen, for example, by substituting the contravariant field tensor and the covariant dual tensor in Cartesian coordinates:

$$F^{\mu\nu} = \begin{pmatrix} 0 & -E_x & -E_y & -E_z \\ E_x & 0 & -B_z & B_y \\ E_y & B_z & 0 & -B_x \\ E_z & -B_y & B_x & 0 \end{pmatrix} \quad {}^*F_{\mu\nu} = \sqrt{-g}\begin{pmatrix} 0 & -B_x & -B_y & -B_z \\ B_x & 0 & -E_z & E_y \\ B_y & E_z & 0 & -E_x \\ B_z & -E_y & E_x & 0 \end{pmatrix}^* \tag{2.20}$$

into (2.19) to yield:

$$R_{\mu\nu} = \tfrac{\kappa}{4\pi} g_{\tau\nu} F^{\tau\sigma} {}^*F_{\mu\sigma} = \tfrac{\kappa}{4\pi} g_{\mu\nu} \sqrt{-g}\left(E_x B_x + E_y B_y + E_z B_z\right) = \tfrac{\kappa}{4\pi} g_{\mu\nu} \sqrt{-g}\, \mathbf{E}\cdot\mathbf{B}. \tag{2.21}$$

Thus, $R_{\mu\nu}$ is seen to be proportional to the scalar $\sqrt{-g}\,\mathbf{E}\cdot\mathbf{B}$ times a $g_{\mu\nu}$ multiplier, and is symmetric because $g_{\mu\nu} = g_{\nu\mu}$. Looking ahead to the following section where we will seek to relate our results to the Euler tensor $T^{\mu\nu} = (\rho_0 + p_0)u^\mu u^\nu + p_0 g^{\mu\nu}$ where $\rho_0$ is the proper energy density of rest matter and $p_0$ is the proper density of energy due to pressure, it is very important to make note of this fact that $R_{\mu\nu} = \mathbf{scalar} \cdot g_{\mu\nu}$ where $\mathbf{scalar} = \frac{\kappa}{4\pi}\sqrt{-g}\mathbf{E}\cdot\mathbf{B}$, i.e., that $R_{\mu\nu}$ is proportional to $g_{\mu\nu}$ times a scalar. This is thus parallel to the pressure term $p_0 g^{\mu\nu}$ in the Euler tensor, where $p_0$ is the scalar. Note also, if we use $*F_{\mu\sigma}F_\nu{}^\sigma = *F_{\nu\sigma}F_\mu{}^\sigma = F_{\mu\sigma}*F_\nu{}^\sigma$ to express this $\mu \Leftrightarrow \nu$ symmetry, that $\mu \Leftrightarrow \nu$ symmetry is equivalent to a symmetry which interchanges $1 \Leftrightarrow *$, i.e., which moves * from $F_{\nu\sigma}$ to $F_\mu{}^\sigma$ and the implied "1" from $F_\mu{}^\sigma$ to $F_{\nu\sigma}$ in (2.19).

---

[*] To see how the determinant $g = \det(g_{\mu\nu})$ arises, consider local geodesic coordinates $g_{\mu\nu} = \eta_{\mu\nu}$. Lower $\varepsilon^{0123}$ by writing $\varepsilon_{0123} = \eta_{00}\eta_{11}\eta_{22}\eta_{33}\varepsilon^{0123} = g\varepsilon^{0123} = -\varepsilon^{0123}$, so that $\varepsilon^{0123}\varepsilon_{0123} = -1$. Generalizing, this yields $\varepsilon_{\mu\nu\sigma\tau} = g\varepsilon^{\mu\nu\sigma\tau}$. Because it is important to normalize covariant and contravariant indexes to 1, we set $\varepsilon^{\delta\eta\mu\sigma}\varepsilon_{\alpha\beta\nu\sigma} = -\delta^{\delta\eta\mu\sigma}{}_{\alpha\beta\nu\sigma}$, and so must finally set $\varepsilon_{0123} = \sqrt{-g}$ and $\varepsilon^{0123} = \dfrac{-1}{\sqrt{-g}}$.

Third, and most importantly, unlike the Maxwell energy tensor which is traceless because the photon has a zero mass, we see that $T^\mu{}_\nu(A)$ has a *non-zero trace*, i.e. a *non-zero rest mass density*, and that the curvature scalar R is therefore also non-zero, as such:

$$T(A) = \Im(A) = \tfrac{1}{\kappa}R = \tfrac{1}{4\pi}*F^{\sigma\tau}F_{\sigma\tau} = \tfrac{1}{\pi}\sqrt{-g}(E_x B_x + E_y B_y + E_z B_z) = \tfrac{1}{\pi}\sqrt{-g}\mathbf{E}\cdot\mathbf{B} \neq 0. \tag{2.22}$$

This means, transforming to a rest frame $u^\mu = (1,0,0,0)$, $T^{0k} = T^{ij} = 0$, that the *rest mass* M(A) = E(A) (in $c^2=1$ units) enclosed within a given volume $V = \int\sqrt{-g}d^3x$ is:

$$M(A) = \int T^0{}_0(A)d^3x = \tfrac{1}{\pi}\int(\sqrt{-g}\mathbf{E}\cdot\mathbf{B})d^3x. \tag{2.23}$$

For an electromagnetic field, the Poynting vector cross product $T^{ok}{}_{Maxwell} = \tfrac{1}{4\pi}(\mathbf{E}\times\mathbf{B})^k$ is of course non-zero, but we generally regard $\mathbf{E}\cdot\mathbf{B} = 0$.[*] That is, in an electromagnetic field, **E** and **B** are generally thought to be orthogonal to one another at any given point in spacetime. Now, turning to the problem of trace matter, it appears that it is a *non-orthogonal* relationship between electric and magnetic fields at any given point in spacetime, that is, a non-zero dot product $\mathbf{E}\cdot\mathbf{B} \neq 0$, which is responsible for a non-zero density of trace matter also existing at that same point in spacetime. We are thus tempted to identify $\Im^\mu{}_\nu(A)$ in (2.15) as the energy tensor used to bring about non-zero trace matter, that is, matter *other than* a pure, traceless electromagnetic field. If this is so, then $\Im^\mu{}_\nu(A)$ may help us better understand the role of gravitation and electromagnetic fields in the structure of trace matter.

Identity (2.4)

Having completed our initial examination of identity (2.5), we now explore identity (2.4). The first term $\tfrac{1}{2}F^{\sigma\tau}(*F_{\tau\sigma;\nu} + *F_{\sigma\nu;\tau} + *F_{\nu\tau;\sigma})$ can be rewritten:

$$\tfrac{1}{2}F^{\mu\sigma}(*F_{\mu\nu;\sigma} + *F_{\nu\sigma;\mu} + *F_{\sigma\mu;\nu})$$
$$= (F^{\mu\sigma}*F_{\nu\sigma} - \tfrac{1}{2}\delta^\mu{}_\nu F^{\sigma\tau}*F_{\sigma\tau})_{;\mu} - *F_{\nu\mu}F^{\sigma\mu}{}_{;\sigma} + *F^{\sigma\tau}F_{\sigma\tau;\nu}. \tag{2.24}$$

so that identity (2.5) becomes:

---

[*] In fact, Maxwell's equations and the Maxwell-Poynting tensor tell us nothing directly about $\mathbf{E}\cdot\mathbf{B}$.

$$\tfrac{1}{2} F^{\mu\sigma} \left( *F_{\mu\nu;\sigma} + *F_{\nu\sigma;\mu} + *F_{\sigma\mu;\nu} \right) + *F_{\nu\mu} F^{\sigma\mu}{}_{;\sigma}$$
$$= \left( F^{\mu\sigma} * F_{\nu\sigma} - \tfrac{1}{2} \delta^{\mu}{}_{\nu} F^{\sigma\tau} * F_{\sigma\tau} \right)_{;\mu} + \tfrac{1}{2} * F^{\sigma\tau} F_{\sigma\tau;\nu} = 0 \quad (2.25)$$

Using this identity to ensure energy conservation, we write:

$$-\kappa T^{\mu}{}_{\nu}(B)_{;\mu} = -\kappa \left( \Im^{\mu}{}_{\nu}(B) + t^{\mu}{}_{\nu}(B) \right)_{;\mu} = \left( R^{\mu}{}_{\nu} - \tfrac{1}{2} \delta^{\mu}{}_{\nu} R \right)_{;\mu} = 0$$
$$= -\tfrac{\kappa}{4\pi} \left[ \tfrac{1}{2} F^{\mu\sigma} \left( *F_{\mu\nu;\sigma} + *F_{\nu\sigma;\mu} + *F_{\sigma\mu;\nu} \right) + *F_{\nu\mu} F^{\sigma\mu}{}_{;\sigma} \right] \quad (2.26)$$
$$= -\tfrac{\kappa}{4\pi} \left[ \left( F^{\mu\sigma} * F_{\nu\sigma} - \tfrac{1}{2} \delta^{\mu}{}_{\nu} F^{\sigma\tau} * F_{\sigma\tau} \right)_{;\mu} + \tfrac{1}{2} * F^{\sigma\tau} F_{\sigma\tau;\nu} \right] = 0$$

which should be contrasted with (2.14). Here, in contrast to (2.15), the integrable term is:

$$\Im^{\mu}{}_{\nu}(B) \equiv \tfrac{1}{4\pi} \left( F^{\mu\sigma} * F_{\nu\sigma} - \tfrac{1}{2} \delta^{\mu}{}_{\nu} F^{\sigma\tau} * F_{\sigma\tau} \right). \quad (2.27)$$

and in contrast to (2.16), the gravitational field term is:

$$\kappa_{\nu}(B) = t^{\mu}{}_{\nu}(B)_{;\mu} = -\tfrac{1}{8\pi} * F^{\sigma\tau} F_{\sigma\tau;\nu} . \quad (2.28)$$

If we now take $\Im^{\mu}{}_{\nu}(B)$ in (2.27) to be a source of matter and so write (contrast (2.17)):

$$-\kappa T^{\mu}{}_{\nu}(B) = -\kappa \Im^{\mu}{}_{\nu}(B) = \left( R^{\mu}{}_{\nu} - \tfrac{1}{2} \delta^{\mu}{}_{\nu} R \right) = -\tfrac{\kappa}{4\pi} \left( F^{\mu\sigma} * F_{\nu\sigma} - \tfrac{1}{2} \delta^{\mu}{}_{\nu} F^{\sigma\tau} * F_{\sigma\tau} \right), \quad (2.29)$$

we find from (2.26) that (contrast (2.18):

$$\kappa_{\nu}(B) = t^{\mu}{}_{\nu}(B)_{;\mu} = \tfrac{1}{8\pi} * F^{\sigma\tau} F_{\sigma\tau;\nu} = 0 . \quad (2.30)$$

Once again, it is encouraging that neither the electric current $J^{\mu}$ nor the magnetic current $P^{\mu}$ vanishes. Further, from (2.29), the trace term, contrast (2.22), is:

$$T(B) = \Im(B) = \tfrac{1}{\kappa} R = -\tfrac{1}{4\pi} F^{\sigma\tau} * F_{\sigma\tau} = -\tfrac{1}{\pi} \sqrt{-g} \left( E_x B_x + E_y B_y + E_z B_z \right) = -\tfrac{1}{\pi} \sqrt{-g} \mathbf{E} \cdot \mathbf{B} \neq 0, \quad (2.31)$$

At rest, contrasting (2.23), this means that the total *trace mass* enclosed within $V = \int \sqrt{-g} d^3 x$ is:

$$M(B) = \int T^0{}_0 d^3 x = -\tfrac{1}{\pi} \int \left( \sqrt{-g} \mathbf{E} \cdot \mathbf{B} \right) d^3 x . \quad (2.32)$$

and so is the *negative* of the mass M(A) in (2.23). Apparently, if $\Im^{\mu}{}_{\nu}(A)$ in (2.15) is the energy tensor which provides positive trace matter, then $\Im^{\mu}{}_{\nu}(B)$ in (2.27) is energy tensor which provides *negative* trace matter, that is, negative energy which moves backward in time to yield positive energy antimatter.

This is further supported by contrasting $\Im^{\mu}{}_{\nu}(B)$ in (2.27) to $\Im^{\mu}{}_{\nu}(A)$ in (2.15). Because we now know that $*F^{\mu\sigma} F_{\nu\sigma} = F^{\mu\sigma} * F_{\nu\sigma}$, see (2.19) and the paragraph thereafter, this means that indeed, $\Im^\mu{}_\nu(A) = -\Im^\mu{}_\nu(B)$. If $\Im^\mu{}_\nu(A)$ represents positive energy, forward temporal matter, then $\Im^\mu{}_\nu(B)$, which derives from the duality rotation $F^{\mu\nu} \to *F^{\mu\nu}$ of identity (2.4) into (2.5), represents negative energy particles of matter moving backwards through time which are equivalent to positive energy antiparticles of matter moving forward in time.[*] Suddenly, we find that duality symmetry appears very much tied not only with chiral symmetry (see [7], section 4), but also with matter and antimatter. So, the same breaking of duality symmetry to reveal electric charges but highly suppress magnetic charges (see [5], equation (8.20)) may also be responsible for the high preponderance in nature, of matter over antimatter.

Mathematically speaking, *irrespective* of the interpretation we give to $\Im^\mu{}_\nu(B)$, given that $*F^{\mu\sigma}F_{\nu\sigma} = F^{\mu\sigma}*F_{\nu\sigma}$, we can contrast the last lines of (2.14) and (2.26) to establish that:

$$\kappa_\nu(A) = t^\mu{}_\nu(A)_{;\mu} = -\tfrac{1}{8\pi}*F^{\sigma\tau}F_{\sigma\tau;\nu} = -\kappa_\nu(B) = -t^\mu{}_\nu(B)_{;\mu} = -\tfrac{1}{8\pi}F^{\sigma\tau}*F_{\sigma\tau;\nu}. \tag{2.33}$$

which means that these terms are also invariant under $1 \Leftrightarrow *$ interchange.

Now, let us return to (2.22), where we first established the connection to non-zero trace matter, and in particular, the term $F^{\sigma\tau}*F_{\sigma\tau} = -4\sqrt{-g}\,\mathbf{E}\cdot\mathbf{B}$. Because this term is a scalar, its covariant derivative is identical to its ordinary derivative, i.e., $(F^{\sigma\tau}*F_{\sigma\tau})_{;\mu} = (F^{\sigma\tau}*F_{\sigma\tau})_{,\mu}$. Making use of (2.33), $*F^{\sigma\tau}F_{\sigma\tau;\nu} = F^{\sigma\tau}*F_{\sigma\tau;\nu}$, which naturally implies $*F^{\sigma\tau}F_{\sigma\tau,\nu} = F^{\sigma\tau}*F_{\sigma\tau,\nu}$ for ordinary derivatives in the $g_{\mu\nu} \to \eta_{\mu\nu}$ limit, we may simplify calculation by replacing some of the covariant derivatives with ordinary derivatives thus:

$$(F^{\sigma\tau}*F_{\sigma\tau})_{;\nu} = (F^{\sigma\tau}*F_{\sigma\tau})_{,\nu} = F^{\sigma\tau}{}_{;\nu}*F_{\sigma\tau} + F^{\sigma\tau}*F_{\sigma\tau;\nu} = *F^{\sigma\tau}F_{\sigma\tau,\nu} + F^{\sigma\tau}*F_{\sigma\tau,\nu}$$
$$= 2F^{\sigma\tau}*F_{\sigma\tau;\nu} = 2F^{\sigma\tau}{}_{;\nu}*F_{\sigma\tau} = 2F^{\sigma\tau}*F_{\sigma\tau,\nu} = 2F^{\sigma\tau}{}_{,\nu}*F_{\sigma\tau} \tag{2.34}$$

Therefore, also using (2.22) and (2.31) together with the energy exchanges expressed by (2.14) and (2.26), we may rewrite (2.16) and (2.28) as follows, employing an *ordinary* derivative of the trace matter scalar $F^{\sigma\tau}*F_{\sigma\tau} = -4\sqrt{-g}\,\mathbf{E}\cdot\mathbf{B}$:

$$\kappa_\nu(A) = t^\mu{}_\nu(A)_{;\mu} = -\Im^\mu{}_\nu(A)_{;\mu} = -\tfrac{1}{8\pi}*F^{\sigma\tau}F_{\sigma\tau;\nu} = -\tfrac{1}{16\pi}(F^{\sigma\tau}*F_{\sigma\tau})_{,\nu} = \tfrac{1}{4\pi}(\sqrt{-g}\,\mathbf{E}\cdot\mathbf{B})_{,\nu}. \tag{2.35}$$

$$\kappa_\nu(B) = t^\mu{}_\nu(B)_{;\mu} = -\Im^\mu{}_\nu(B)_{;\mu} = \tfrac{1}{8\pi}F^{\sigma\tau}*F_{\sigma\tau;\nu} = \tfrac{1}{16\pi}(F^{\sigma\tau}*F_{\sigma\tau})_{,\nu} = -\tfrac{1}{4\pi}(\sqrt{-g}\,\mathbf{E}\cdot\mathbf{B})_{,\nu}. \tag{2.36}$$

These results may be consolidated to write:

---

[*] In interpreting the negative energy solutions in this way, we follow suit with the Feynman-Stückelberg interpretation given to the negative energy solutions of the relativistic Schrödinger (Klein-Gordon) and Dirac equations. See, e.g., [9], Chapter 3, for a good discussion of this. From here, we shall refer to positive energy forward temporal matter simply as "matter" and negative energy backward temporal matter as positive energy forward temporal "antimatter."

$$\kappa_\nu(A) = t^\mu{}_\nu(A)_{;\mu} = -\mathfrak{I}^\mu{}_\nu(A)_{;\mu} = -\kappa_\nu(B) = -t^\mu{}_\nu(B)_{;\mu} = \mathfrak{I}^\mu{}_\nu(B)_{;\mu}$$

$$= \tfrac{1}{4\pi}\left(\sqrt{-g}\mathbf{E}\cdot\mathbf{B}\right)_{,\nu} = \tfrac{1}{4\pi}\left(\frac{\partial\left(\sqrt{-g}\mathbf{E}\cdot\mathbf{B}\right)}{\partial x^0}\quad \frac{\partial\left(\sqrt{-g}\mathbf{E}\cdot\mathbf{B}\right)}{\partial x^1}\quad \frac{\partial\left(\sqrt{-g}\mathbf{E}\cdot\mathbf{B}\right)}{\partial x^2}\quad \frac{\partial\left(\sqrt{-g}\mathbf{E}\cdot\mathbf{B}\right)}{\partial x^3}\right). \quad (2.37)$$

$$= \tfrac{1}{4}(T(A))_{,\nu} = \tfrac{1}{4}\left(\frac{\partial T(A)}{\partial x^0}\quad \frac{\partial T(A)}{\partial x^1}\quad \frac{\partial T(A)}{\partial x^2}\quad \frac{\partial T(A)}{\partial x^3}\right)$$

Now, let us consider what it means when $\kappa_\nu = 0$, and when $\kappa_\nu \neq 0$.

Discussion

Referring to (2.37), for a region of spacetime where $\kappa_\nu(A) = 0$, equation (2.18), or $\kappa_\nu(B) = 0$, equation (2.30), this means that the trace density $T(A)$ of positive energy forward temporal matter ("matter"), or $T(B)$ of negative energy backward temporal matter ("antimatter"), is constant, i.e., *uniformly-distributed*. Simultaneously, $\kappa_\nu(A) = 0$, $\kappa_\nu(B) = 0$ describes a circumstance where *no energy is exchanged* between matter (or antimatter) and the gravitational field, i.e., a state of energy equilibrium. More to the point, equation (2.37), *when $\kappa_\nu = 0$*, says: "Uniform Density of Matter Over a Finite Space = Energy Equilibrium Within the Finite Space." Thus, $\kappa_\nu(A) = 0$, $\kappa_\nu(B) = 0$, describes a state where trace matter (which may be non-zero) is *uniformly-distributed* and in *material equilibrium with the gravitational field*.

This is in contrast to $\kappa_\nu = 0$ in Maxwell's electrodynamics (see [1], equation (65) and thereafter, as well as above, equations (1.6), (2.8) or following (2.11)), which causes the sources of charge $J^\mu$, $P^\mu$ to become zero, and where the only matter which exists is the traceless matter (photons) of the electromagnetic field as well as the energy of the pure gravitational field. That is, $\kappa_\nu = 0$ in [1] describes a region of spacetime filled with nothing but electromagnetic fields, i.e., photons (light), and gravitational fields. In contrast, $\kappa_\nu(A) = 0$, $\kappa_\nu(B) = 0$ describes a region of spacetime where sources of charge $J^\mu$, $P^\mu$ can and do exist, and where trace matter – which also can and does exist – is uniformly-distributed and in energy equilibrium. Identities (2.2) and (2.3) form the basis for describing light as a material phenomenon. Identities (2.4) and (2.5) form the basis for the describing trace matter itself.

If we take $\kappa_\nu(A) = 0$, $\kappa_\nu(B) = 0$ to describe a uniform density of matter which goes hand in hand with energy equilibrium, then we can ask about the dynamical consequences when $\kappa_\nu(A) \neq 0$, $\kappa_\nu(B) \neq 0$. Here, energy is distributed *non-uniformly*, because the 4-gradient in (2.37) has one or more non-zero components. At the same time, energy is being dynamically exchanged between $\mathfrak{I}^\mu{}_\nu(A)$ and $t^\mu{}_\nu(A)$, and / or between $\mathfrak{I}^\mu{}_\nu(B)$ and $t^\mu{}_\nu(B)$, that is, between matter and the gravitational field. This may be envisioned, for example, to describe a non-uniform distribution of molecular gas energy in an enclosed space, which, over time, will readjust its density toward a uniform distribution. This may be envisioned to describe a solid, thermally-conductive material body which is warmer in one region and colder in another, and which will therefore transfer heat from the warmer region to the colder region until the thermal energy is fully balanced. This may be envisioned to describe an electrically-conductive material in which an equal number of positive and negative charges are non-uniformly distributed, wherein these charges will eventually rearrange themselves into a substantially neutral, stable state throughout the material. In all cases, wherever $\kappa_\nu(A) \neq 0$, $\kappa_\nu(B) \neq 0$, matter will exchange energy with its gravitational field, and these exchanges will continue apace, with $\kappa_\nu(A)$, $\kappa_\nu(B)$ → 0, until the energy is uniformly-distributed. Then, when $\kappa_\nu(A) = 0$, $\kappa_\nu(B) = 0$, all matter is in uniform distribution because the four-gradient has vanished and no more energy exchanges occur because the same vanishing four-gradient also zeros the exchange of energy between matter and the gravitational field. In short, the system has now attained equilibrium. By virtue of $t^\mu{}_\nu(A)$ and $t^\mu{}_\nu(B)$ being involved in these exchanges, the broad range of phenomena where non-uniform distributions of energy seek a uniform equilibrium through energy exchanges is fully understood as *gravitational* phenomena involving matter exchanging energy with the gravitational field until the equilibrium is reached.

It is also worth considering this in terms of the *conversion* of energy from one form to another. Let us postulate that $t^\mu{}_\nu(A)_{;\mu} = -\frac{1}{8\pi} F^{\sigma\tau} * F_{\sigma\tau;\nu} \cong 0$ at a point in spacetime, but is not *exactly* equal to zero. From (2.14), $\mathfrak{I}^\mu{}_\nu(A) = -\frac{1}{4\pi}\left(* F^{\mu\sigma} F_{\nu\sigma} - \frac{1}{2}\delta^\mu{}_\nu * F^{\sigma\tau} F_{\sigma\tau}\right) \cong \left(R^\mu{}_\nu - \frac{1}{2}\delta^\mu{}_\nu R\right)$ as well, because $\frac{\kappa}{4\pi}\left(* F^{\mu\sigma} F_{\nu\sigma} - \frac{1}{2}\delta^\mu{}_\nu * F^{\sigma\tau} F_{\sigma\tau}\right)_{;\mu} \cong 0$. If this is the case, then the energy tensor $T^\mu{}_{\nu\,matter}$ of *matter in some form*, defined as $T^\mu{}_{\nu\,matter} \equiv \left(R^\mu{}_\nu - \frac{1}{2}\delta^\mu{}_\nu R\right)$, is not quite equal to $\mathfrak{I}^\mu{}_\nu(A)$, but is only *approximately* equal to $\mathfrak{I}^\mu{}_\nu(A)$, i.e., $T^\mu{}_{\nu\,matter} = \mathfrak{I}^\mu{}_\nu(A) + \delta T^\mu{}_\nu$, where by virtue of our postulate, $\delta T^\mu{}_\nu$ is the energy tensor for matter in some form *other than the form of* $\mathfrak{I}^\mu{}_\nu(A)$. As we shall explore in the following section, there are a variety of energy tensors which describe energy in different forms, whether due to electromagnetic fields or trace matter or pressure or motion, etc. At any given point in spacetime, there can of course be energy due to rest mass, and pressure, and heat, and kinetic and potential energies, and electromagnetic fields, to name a few, so it should not be surprising to find a composition of energy tensors for distinct forms of energy at a given point in spacetime. But, what is of interest here is that this coincides with $t^\mu{}_\nu(A)_{;\mu} = -\frac{1}{8\pi} F^{\sigma\tau} * F_{\sigma\tau;\nu} \cong 0$. Because, if $t^\mu{}_\nu(A)_{;\mu} = -\frac{1}{8\pi} F^{\sigma\tau} * F_{\sigma\tau;\nu} = 0$ is regarded as the equilibrium state where there is no longer an exchange of energy between matter and gravitation, then energy exchange between matter and gravitation when $t^\mu{}_\nu(A)_{;\mu} = -\frac{1}{8\pi} F^{\sigma\tau} * F_{\sigma\tau;\nu} \cong 0$ is equivalent to the exchange of energy between $\mathfrak{I}^\mu{}_\nu(A)$ and $\delta T^\mu{}_\nu$, and is thus another way of saying that energy is being *converted* from one material form to another.

Thus, not only can we regard the tendency of a system of energy toward equilibrium to be a gravitational effect, but we can regard the conversion of matter from one material form to another – for example, electrical to heat, mechanical to electrical, nuclear to kinetic, etc. – to be described, fundamentally, by $t^\mu{}_\nu(A)_{;\mu}$, that is, as a phenomenon resulting from the exchange of energy between matter and the gravitation field. Thus, in seeking an equilibrium state – which is now seen to be a gravitational phenomenon – one of the things that energy will do is convert from one material form to another. The conversion itself thus becomes a gravitational phenomenon that results from the tendency of energy to seek the equilibrium state, $t^\mu{}_{\nu;\mu} = 0$.

Finally, if we do indeed regard $\mathfrak{I}^\mu{}_\nu(B) = -\mathfrak{I}^\mu{}_\nu(A)$ as a tensor for negative energy moving backwards through time, then by taking these as equivalent to positive energy antiparticles moving forward in time, it is no longer necessary to maintain $\mathfrak{I}^\mu{}_\nu(A)$ and $\mathfrak{I}^\mu{}_\nu(B)$ as separate tensors, and in general, we don't need to make an explicit A = +energy, B = -energy distinction any longer. Thus, the tensor for all positive trace energy, forward temporal, matter and antimatter, from (2.15), is:

$$\Im_{\mu\nu} = -\tfrac{1}{4\pi}\left(*F_\mu{}^\sigma F_{\nu\sigma} - \tfrac{1}{2} g_{\mu\nu} *F^{\sigma\tau} F_{\sigma\tau}\right) \tag{2.38}$$

while the exchange of energy between this trace matter and the gravitational field, from (2.35), is described by:

$$\kappa_\nu = t^\mu{}_{\nu;\mu} = \tfrac{1}{16\pi}\left(F^{\sigma\tau} * F_{\sigma\tau}\right)_{,\nu}. \tag{2.39}$$

To specify the geometrodynamics through which any system of energy seeks its equilibrium state $\kappa_\nu = 0$, we actually set $\kappa_\nu = 0$ in the Einstein equation, and therefore equate the geometric term $R^\mu{}_\nu - \tfrac{1}{2}\delta^\mu{}_\nu R$ to the energy tensor of *matter only*, for whatever material form of energy is represented by the energy tensor. This implicitly incorporates gravitational effects into the geometric term $R^\mu{}_\nu - \tfrac{1}{2}\delta^\mu{}_\nu R$ as well. For the trace matter tensor $\Im_{\mu\nu}$, including $R^\mu{}_\nu$ and its trace $R$ from (2.21), we write this as:

$$-\kappa\Im_{\mu\nu} = R_{\mu\nu} - \tfrac{1}{2}g_{\mu\nu}R = -\tfrac{\kappa}{4\pi} g_{\mu\nu}\sqrt{-g}\,\mathbf{E}\cdot\mathbf{B}. \tag{2.40}$$

The Einstein equation above, as applied to trace matter, then specifies in exact, non-linear terms, the dynamical preference of any system of energy to redistribute itself by a geodesically-based flow of energy, until an equilibrium state of uniform energy density is achieved. Setting $\kappa_\nu = 0$ for a given energy tensor $T^\mu{}_{\nu\,matter}$ (whether $\Im_{\mu\nu}$ or some other energy tensor, as explored in the next section), and simultaneously setting that energy tensor $T^\mu{}_{\nu\,matter}$ to $R^\mu{}_\nu - \tfrac{1}{2}\delta^\mu{}_\nu R$, establishes $T^\mu{}_{\nu\,matter}$ as the "equilibrium" state and can be thought of as a "spring" which "winds up" the Einstein equation to seek out $T^\mu{}_{\nu\,matter} = R^\mu{}_\nu - \tfrac{1}{2}\delta^\mu{}_\nu R$ as the equilibrium state, so that matter is converted and redistributed as necessary, until that equilibrium state is achieved. That is, $T^\mu{}_{\nu\,matter} = R^\mu{}_\nu - \tfrac{1}{2}\delta^\mu{}_\nu R$ biases the Einstein equation so that in the dynamical course of geodesic events, matter will convert and redistribute as needed, to reach the $\kappa_\nu = 0$ state for the particular $T^\mu{}_{\nu\,matter}$ specified.

Now, we examine several energy tensors which may be derived from equation (2.40).

## 3. Derivation of Energy Tensors for Matter Flux, Pressure Flux, and Stationary Pressure, as a Function of Electromagnetic Fields

Let us now assemble together, several different formulations of the energy tensor which will be important in the discussion to follow. First, let us consider the Euler tensor mentioned earlier, just after equation (2.21):

$$T^{\mu\nu}{}_{Euler} = (\rho_0 + p_0)u^\mu u^\nu + p_0 g^{\mu\nu}. \tag{3.1}$$

Next, let us consider the Maxwell – Poynting tensor of (1.5), in contravariant form:

$$T^{\mu\nu}{}_{Maxwell} = -\tfrac{1}{4\pi}\left[F^{u\sigma}F^{v}{}_{\sigma} - \tfrac{1}{4}g^{\mu\nu}F^{\tau\sigma}F_{\tau\sigma}\right], \tag{3.2}$$

Finally, let us consider the trace matter tensor in (2.38), (2.40), written as:

$$\Im^{\mu\nu} = -\tfrac{1}{4\pi}\left(*F^{\mu\sigma}F^{v}{}_{\sigma} - \tfrac{1}{2}g^{\mu\nu}*F^{\sigma\tau}F_{\sigma\tau}\right) = \tfrac{1}{4\pi}g^{\mu\nu}\sqrt{-g}\mathbf{E}\cdot\mathbf{B} \tag{3.3}$$

Our interest here, is to derive an energy tensor for matter in flux, $T^{\mu\nu}{}_{Matter\ Flux} \equiv \rho_0 u^\mu u^\nu$, stationary pressure, $T^{\mu\nu}{}_{Stationary\ Pressure} \equiv p_0 g^{\mu\nu}$, and pressure energy flux, $T^{\mu\nu}{}_{Pressure\ Flux} = p_0 u^\mu u^\nu$, as separate and distinct tensors, *all in terms of the electromagnetic field tensor $F^{uv}$*.

Let us first consider three distinct characteristics of an energy tensor. First, whether its trace is zero or non-zero. Second, whether it contains a term which is proportional to the metric tensor times a scalar, that is, a term of form $g^{\mu\nu} \cdot \mathbf{scalar}$. Third, whether it contains a term which *cannot* be expressed as $g^{\mu\nu} \cdot \mathbf{scalar}$, and which therefore may be expressed as $u^\mu u^\nu \cdot \mathbf{scalar}$, that is, as a net kinetic flux.

Equation (3.1) for $T^{\mu\nu}{}_{Euler}$ has all three of these characteristics. First, its trace is:

$$T_{Euler} = (\rho_0 + p_0) + 4p_0 = \rho_0 + 5p_0. \tag{3.4}$$

The fifth $p_0$ in the $5p_0$ term arises on account of the fact that any pressure applied to a rest energy density $\rho_0$, will increase the rest energy of that rest energy density, and so will add to the energy flux when that pressurized energy density is also set into motion. Second, $T^{\mu\nu}{}_{Euler}$ contains a term of the form $g^{\mu\nu} \cdot \mathbf{scalar}$, namely, $p_0 g^{\mu\nu}$. Third, $T^{\mu\nu}{}_{Euler}$ contains a term of the form $(\rho_0 + p_0) u^\mu u^\nu$ which is *not* proportional to $g^{\mu\nu}$.

The Maxwell tensor has only two these three characteristics. Its trace, of course, vanishes (see (1.9)):

$$T_{Maxwell} \equiv -\tfrac{1}{4\pi}\left[F^{\tau\sigma}F_{\tau\sigma} - F^{\tau\sigma}F_{\tau\sigma}\right] = 0, \tag{3.5}$$

and so that condition does not obtain. But, it does contain a term of the form $g^{\mu\nu} \cdot \mathbf{scalar}$, where $\mathbf{scalar} = \tfrac{1}{16\pi}F^{\tau\sigma}F_{\tau\sigma}$. And, it does contain a term which *cannot* be expressed as $g^{\mu\nu} \cdot \mathbf{scalar}$, namely, $-\tfrac{1}{4\pi}F^{u\sigma}F^{v}{}_{\sigma}$ (substitute explicit components from (2.20) to see this).

The trace matter tensor (3.3), also has only two of these three characteristics. It does, as we have seen, have a non-zero trace, see (2.22):

$$\Im = \tfrac{1}{4\pi}*F^{\sigma\tau}F_{\sigma\tau} = \tfrac{1}{\pi}\sqrt{-g}\mathbf{E}\cdot\mathbf{B} \tag{3.6}$$

It also has a term of the form of the form $g^{\mu\nu} \cdot \mathbf{scalar}$, namely, $\mathbf{scalar} = \tfrac{1}{4\pi}\sqrt{-g}\mathbf{E}\cdot\mathbf{B}$. But, it does *not* contain a term which *cannot* be expressed as $g^{\mu\nu} \cdot \mathbf{scalar}$, and thus related to the flux of material energy, $(\rho_0 + p_0)u^\mu u^\nu$. In short, examining the structural characteristics of the energy tensors (3.2) and (3.3), it is clear that neither one of these, *separately*, can be properly set to the Euler tensor (3.1). The Maxwell tensor lacks the necessary trace, and $\mathfrak{I}^{\mu\nu}$ does not contain a term that can be related to the motion term $u^\mu u^\nu$.

Let us, therefore, consider *combining* (3.2) and (3.3) to form the Euler tensor. That is, let us define a new tensor which is the *sum* of (3.2) and (3.3),

$$T^{\mu\nu}{}_{Combined} = T^{\mu\nu}{}_{Maxwell} + \mathfrak{I}^{\mu\nu} = -\tfrac{1}{4\pi}\left[F^{u\sigma}F^\nu{}_\sigma - \tfrac{1}{4}g^{\mu\nu}F^{\tau\sigma}F_{\tau\sigma}\right] - \tfrac{1}{4\pi}\left(*F^{\mu\sigma}F^\nu{}_\sigma - \tfrac{1}{2}g^{\mu\nu}*F^{\sigma\tau}F_{\sigma\tau}\right), \quad (3.7)$$

and let us equate this to the Euler Tensor (3.1) to yield:

$$\begin{aligned}T^{\mu\nu}{}_{Euler} &= T^{\mu\nu}{}_{Combined} = T^{\mu\nu}{}_{Maxwell} + \mathfrak{I}^{\mu\nu} = (\rho_0 + p_0)u^\mu u^\nu + p_0 g^{\mu\nu}\\ &= -\tfrac{1}{4\pi}\left[F^{u\sigma}F^\nu{}_\sigma + *F^{\mu\sigma}F^\nu{}_\sigma - \tfrac{1}{4}g^{\mu\nu}F^{\tau\sigma}F_{\tau\sigma} - \tfrac{1}{2}g^{\mu\nu}*F^{\sigma\tau}F_{\sigma\tau}\right]\end{aligned}, \quad (3.8)$$

The above combination of terms containing the electromagnetic field tensor $F^{u\nu}$ now has all three of the characteristics we seek: non-zero trace, terms of the form $g^{\mu\nu} \cdot$ **scalar**, and a term which is *not* of the form $g^{\mu\nu} \cdot$ **scalar**. From here, let us separate the terms which are of the form $g^{\mu\nu} \cdot$ **scalar**, from those which are not. The latter two terms, $-\tfrac{1}{2}g^{\mu\nu}*F^{\sigma\tau}F_{\sigma\tau} - \tfrac{1}{4}g^{\mu\nu}F^{\tau\sigma}F_{\tau\sigma}$, manifestly are of the form $g^{\mu\nu} \cdot$ **scalar**. But what about the other two terms? The term $*F^{\mu\sigma}F^\nu{}_\sigma$ is recognizable from the Ricci tensor of (2.21), rewritten as:

$$R^{\mu\nu} = \tfrac{\kappa}{4\pi}*F^{\mu\sigma}F^\nu{}_\sigma = \tfrac{\kappa}{4\pi}g^{\mu\nu}\sqrt{-g}\mathbf{E}\cdot\mathbf{B}, \quad (3.9)$$

and so is *also* of the form $g^{\mu\nu} \cdot$ **scalar**. Finally, the term $F^{u\sigma}F^\nu{}_\sigma$, which is the first term in the Maxwell tensor (3.2), is *not* capable of being reduced to the form $g^{\mu\nu} \cdot$ **scalar**, as just noted.

So, based on the proportionality, or lack thereof, to $g^{\mu\nu}$, we separate (3.8) into two separate expressions:

$$(\rho_0 + p_0)u^\mu u^\nu = -\tfrac{1}{4\pi}F^{u\sigma}F^\nu{}_\sigma, \quad (3.10)$$

and:

$$p_0 g^{\mu\nu} = -\tfrac{1}{4\pi}\left[*F^{\mu\sigma}F^\nu{}_\sigma - \tfrac{1}{4}g^{\mu\nu}F^{\tau\sigma}F_{\tau\sigma} - \tfrac{1}{2}g^{\mu\nu}*F^{\sigma\tau}F_{\sigma\tau}\right]. \quad (3.11)$$

We then take the trace of each of the above to yield:

$$\rho_0 + p_0 = -\tfrac{1}{4\pi}F^{\sigma\tau}F_{\sigma\tau}, \quad (3.12)$$

and:

$$p_0 = \tfrac{1}{4\pi}\left[\tfrac{1}{4}F^{\sigma\tau}F_{\sigma\tau} + \tfrac{1}{4}*F^{\sigma\tau}F_{\sigma\tau}\right]. \quad (3.13)$$

Subtracting (3.13) from (3.12) then yields:

$$\rho_0 = -\frac{1}{4\pi}\left[\frac{5}{4}F^{\sigma\tau}F_{\sigma\tau} + \frac{1}{4}*F^{\sigma\tau}F_{\sigma\tau}\right]. \tag{3.14}$$

Now, we return to (3.10), and substitute the results from (3.13) and (3.14) to yield:

$$\begin{aligned}(\rho_0 + p_0)u^\mu u^\nu &= -\frac{1}{4\pi}\left[\frac{5}{4}F^{\tau\sigma}F_{\tau\sigma} + \frac{1}{4}*F^{\sigma\tau}F_{\sigma\tau}\right]u^\mu u^\nu + \frac{1}{4\pi}\left[\frac{1}{4}F^{\tau\sigma}F_{\tau\sigma} + \frac{1}{4}*F^{\sigma\tau}F_{\sigma\tau}\right]u^\mu u^\nu \\ &= -\frac{1}{4\pi}F^{\mu\sigma}F^\nu{}_\sigma = -\frac{1}{4\pi}\left[\frac{5}{4}F^{\mu\sigma}F^\nu{}_\sigma + \frac{1}{4}*F^{\mu\sigma}F^\nu{}_\sigma\right] + \frac{1}{4\pi}\left[\frac{1}{4}F^{\mu\sigma}F^\nu{}_\sigma + \frac{1}{4}*F^{\mu\sigma}F^\nu{}_\sigma\right]\end{aligned} \tag{3.15}$$

where in the final term, we have constructed a straightforward identity from $-\frac{1}{4\pi}F^{\mu\sigma}F^\nu{}_\sigma$. Thus, from the above, we may identify the separate tensors:

$$T^{\mu\nu}{}_{Matter\ Flux} = \rho_0 u^\mu u^\nu = -\frac{1}{4\pi}\left[\frac{5}{4}F^{\mu\sigma}F^\nu{}_\sigma + \frac{1}{4}*F^{\mu\sigma}F^\nu{}_\sigma\right], \text{ and} \tag{3.17}$$

$$T^{\mu\nu}{}_{Pressure\ Flux} = p_0 u^\mu u^\nu = \frac{1}{4\pi}\left[\frac{1}{4}F^{\mu\sigma}F^\nu{}_\sigma + \frac{1}{4}*F^{\mu\sigma}F^\nu{}_\sigma\right], \tag{3.18}$$

together with (3.11) now written as:

$$T^{\mu\nu}{}_{Stationary\ Pressure} = p_0 g^{\mu\nu} = -\frac{1}{4\pi}\left[*F^{\mu\sigma}F^\nu{}_\sigma - \frac{1}{4}g^{\mu\nu}F^{\tau\sigma}F_{\tau\sigma} - \frac{1}{2}g^{\mu\nu}*F^{\sigma\tau}F_{\sigma\tau}\right]. \tag{3.19}$$

The overall Euler tensor (3.8), which is the sum of all three tensors above, is written as:

$$T^{\mu\nu}{}_{Euler} = (\rho_0 + p_0)u^\mu u^\nu + p_0 g^{\mu\nu} = -\frac{1}{4\pi}\left[F^{\mu\sigma}F^\nu{}_\sigma + *F^{\mu\sigma}F^\nu{}_\sigma - \frac{1}{4}g^{\mu\nu}F^{\tau\sigma}F_{\tau\sigma} - \frac{1}{2}g^{\mu\nu}*F^{\sigma\tau}F_{\sigma\tau}\right]. \tag{3.20}$$

The respective trace relationships are (contrast with (3.13) and (3.14)):

$$T_{Matter\ Flux} = \rho_0 = -\frac{1}{4\pi}\left[\frac{5}{4}F^{\sigma\tau}F_{\sigma\tau} + \frac{1}{4}*F^{\sigma\tau}F_{\sigma\tau}\right], \tag{3.21}$$

$$T_{Pressure\ Flux} = p_0 = \frac{1}{4\pi}\left[\frac{1}{4}*F^{\sigma\tau}F_{\sigma\tau} + \frac{1}{4}F^{\sigma\tau}F_{\sigma\tau}\right], \tag{3.22}$$

$$T_{Stationary\ Pressure} = 4p_0 = \frac{1}{4\pi}\left[*F^{\sigma\tau}F_{\sigma\tau} + F^{\tau\sigma}F_{\tau\sigma}\right], \text{ and} \tag{3.23}$$

$$T_{Euler} = \rho_0 + 5p_0 = \frac{1}{4\pi}*F^{\sigma\tau}F_{\sigma\tau} = \frac{1}{\pi}\sqrt{-g}\,\mathbf{E}\cdot\mathbf{B}. \tag{3.24}$$

It will also be helpful to refer to the total kinetic flux defined as:

$$T^{\mu\nu}{}_{Flux} = (\rho_0 + p_0)u^\mu u^\nu = -\frac{1}{4\pi}F^{\mu\sigma}F^\nu{}_\sigma. \tag{3.25}$$

from (3.17) and (3.18). The trace of this is:

$$T_{Flux} = (\rho_0 + p_0) = -\frac{1}{4\pi}F^{\sigma\tau}F_{\sigma\tau}. \tag{3.26}$$

These are all constructed solely out of various combinations of the electromagnetic field strength tensor $F^{\mu\nu}$, and they all have non-zero trace. Equations (3.12), $\rho_0 + p_0 = -\frac{1}{4\pi}F^{\sigma\tau}F_{\sigma\tau}$ and (3.24), $\rho_0 + 5p_0 = \frac{1}{4\pi}*F^{\sigma\tau}F_{\sigma\tau}$, are equations which "subsist" between $\rho_0$ and $p_0$, see [1], start of §19. Interestingly, by decomposing Euler's equation into the separate equations (3.10), (3.11) based on whether or not the $F^{\mu\nu}$ terms are equivalent to $g^{\mu\nu} \cdot$ **scalar**, we find now that the energy tensors (3.17) and (3.18) do contain a $g^{\mu\nu} \cdot$ **scalar** term (the one involving $*F^{\sigma\tau}F_{\sigma\tau}$), as well as a term that is not expressible as $g^{\mu\nu} \cdot$ **scalar** (the one involving $F^{\sigma\tau}F_{\sigma\tau}$). This "mixing" of terms is a direct result of the $(\rho_0 + p_0)u^\mu u^\nu$ term in Euler's equation (3.1), that is, of the fact that pressure itself creates energy, and when in flux, thus adds to the flow of rest energy. The final result, $T_{Euler} = \frac{1}{\pi}\sqrt{-g}\mathbf{E}\cdot\mathbf{B}$, explains the trace in (2.22) as the trace of the Euler energy tensor. It is of course of great interest that the Euler tensor itself, $T^{\mu\nu}{}_{Euler} = T^{\mu\nu}{}_{Maxwell} + \Im^{\mu\nu}$, see (3.8), also contains Maxwell's tensor as well as the $\Im^{\mu\nu}$ developed here. That latter tensor, $\Im^{\mu\nu} = T^{\mu\nu}{}_{Euler} - T^{\mu\nu}{}_{Maxwell}$, also employed for example in (2.40), can therefore be regarded as Euler's tensor in the absence of any electromagnetic field energy.

One thing that has not been considered explicitly in any of these energy tensors is temperature. While one may be tempted to add a term $kT$ to the pressure term in the Euler tensor to represent thermal energy, we shall resist the temptation to do so here. In particular, because Hawking has demonstrated the emergence of a blackbody spectrum from a black hole event horizon in which surface area behaves like entropy and surface gravity like temperature, there appear to be good reasons to *not* impose any of the laws of thermodynamics on the energy tensors, but rather, to await their natural emergence from solutions to the Einstein equations.

The development of the various energy tensors elaborated above, integrates into an organic whole, Einstein's consideration of Material Phenomena in section D of [1], both §19 and §20, and resolves some of the fundamental questions which Einstein sought to address in [6].

## 4. Third-Order Conditions on the Metric Tensor, Using the Gravitational Energy Components

Let us return now take a further look at the role of the gravitational components $\kappa_\nu \equiv t^\mu{}_{\nu;\mu}$ and the process by which these components vanish, $\kappa_\nu = 0$. We will seek to apply this analysis to all of the energy tensors developed in section 3.

First, we rewrite equation (2.7), now *without* the Bianchi identity $\left(R^\mu{}_\nu - \frac{1}{2}\delta^\mu{}_\nu R\right)_{;\mu}$:

$$
\begin{aligned}
-\kappa T^\mu{}_{\nu;\mu} &= -\kappa\left(T^\mu{}_{\nu\,Maxwell} + t^\mu{}_{\nu\,Maxwell}\right)_{;\mu} = -\kappa T^\mu{}_{\nu\,Maxwell;\mu} - \kappa\cdot\kappa_{\nu\,Maxwell} \\
&= \tfrac{\kappa}{4\pi}\left[\tfrac{1}{2}F^{\mu\sigma}\left(F_{\mu\nu;\sigma} + F_{\nu\sigma;\mu} + F_{\sigma\mu;\nu}\right) - *F_{\nu\mu}*F^{\sigma\mu}{}_{;\sigma}\right] \\
&= \tfrac{\kappa}{4\pi}\left[\left(F^{\mu\sigma}F_{\nu\sigma} - \tfrac{1}{4}\delta^\mu{}_\nu F^{\tau\sigma}F_{\tau\sigma}\right)_{;\mu} - F_{\nu\mu}F^{\sigma\mu}{}_{;\sigma} - *F_{\nu\mu}*F^{\sigma\mu}{}_{;\sigma}\right] = 0
\end{aligned}
\qquad (4.1)
$$

where we have specifically labeled $\kappa_\nu$ as $\kappa_{\nu\,Maxwell}$ in recognition of these being the gravitational energy components for the Maxwell tensor, and in recognition of the fact that the $\kappa_\nu$ may be of a different form for different energy tensors, i.e., for the energy tensors of different *forms* of matter, for example, $\kappa_\nu(A) = \tfrac{1}{8\pi}F^{\sigma\tau}*F_{\sigma\tau;\nu}$ in (2.16) which is associated with $\Im^{\mu\nu}$.

Now, we repeat the step following (2.7) of specifically identifying the Maxwell tensor (1.5), *and associating this with* $R^\mu{}_\nu - \frac{1}{2}\delta^\mu{}_\nu R$, according to:

$$R^\mu{}_\nu - \tfrac{1}{2}\delta^\mu{}_\nu R = -\kappa T^\mu{}_{\nu\,Maxwell} = \tfrac{\kappa}{4\pi}\left(F^{u\sigma}F_{\nu\sigma} - \tfrac{1}{4}\delta^\mu{}_\nu F^{\tau\sigma}F_{\tau\sigma}\right), \tag{4.2}$$

which because of identity (2.2), $\tfrac{1}{2}F^{\mu\sigma}\left(F_{\mu\nu;\sigma} + F_{\nu\sigma;\mu} + F_{\sigma\mu;\nu}\right) - *F_{\nu\mu}*F^{\sigma\mu}{}_{;\sigma} = 0$, immediately introduces the concurrent association:

$$\kappa_{\nu\,Maxwell} = \tfrac{1}{4\pi}\left(F_{\nu\mu}F^{\sigma\mu}{}_{;\sigma} + *F_{\nu\mu}*F^{\sigma\mu}{}_{;\sigma}\right). \tag{4.3}$$

By making the association in (4.2) of $T^\mu{}_{\nu\,Maxwell}$ to $R^\mu{}_\nu - \tfrac{1}{2}\delta^\mu{}_\nu R$, we are specifically choosing to associate the Maxwell tensor, which describes the energy in an electromagnetic field, with "matter." But, if the material phenomena we are choosing to examine in any given circumstance is something other than an electromagnetic field, for example, an adiabatic fluid based on the Euler tensor (3.24), or any of the other energy tensors reviewed in the previous section, or any other energy tensors which one may construct, then we will associate a *different* combination of fields with $R^\mu{}_\nu - \tfrac{1}{2}\delta^\mu{}_\nu R$. To the point: we apparently have the freedom to associate with $R^\mu{}_\nu - \tfrac{1}{2}\delta^\mu{}_\nu R$, *any* energy tensor or combination of energy tensors that we choose. The particular association we choose will depend on the material phenomena to be examined. And, the particular association we choose, by virtue of the identities (2.2) through (2.5), will also serve to fix the associated $\kappa_\nu$, which will also differ depending on the material phenomenon we are considering. The particular $T^\mu{}_{\nu\,Matter}$ which we set to $R^\mu{}_\nu - \tfrac{1}{2}\delta^\mu{}_\nu R$, establishes the equilibrium state toward which the Einstein equation will "gravitate" through geodesically-based redistributions and conversions of energy.

Now, let's take a careful look at what happens when we take the contracted covariant derivative of each side of (4.2), to yield:

$$-\kappa T^\mu{}_{\nu\,Maxwell;\mu} = \tfrac{\kappa}{4\pi}\left(F^{u\sigma}F_{\nu\sigma} - \tfrac{1}{4}\delta^\mu{}_\nu F^{\tau\sigma}F_{\tau\sigma}\right)_{;\mu} = \left(R^\mu{}_\nu - \tfrac{1}{2}\delta^\mu{}_\nu R\right)_{;\mu} = 0. \tag{4.4}$$

Equality to zero in (4.4) comes about from the Bianchi identity. But, in (4.1), the equality to zero comes about from $\tfrac{1}{2}F^{\mu\sigma}\left(F_{\mu\nu;\sigma} + F_{\nu\sigma;\mu} + F_{\sigma\mu;\nu}\right) - *F_{\nu\mu}*F^{\sigma\mu}{}_{;\sigma} = 0$, identity (2.2). If one carefully contrasts (4.1) and (4.4), it should be clear that the zero in (4.1) and the zero in (4.4) are *two different zeros*. In fact, by combining (4.1) and (4.4), and employing the identity (2.2), we find that by virtue of these two independent zeros:

$$\left(R^\mu{}_\nu - \tfrac{1}{2}\delta^\mu{}_\nu R\right)_{;\mu} = -\kappa T^\mu{}_{\nu\,Maxwell;\mu} = \kappa \cdot \kappa_{\nu\,Maxwell} = \tfrac{\kappa}{4\pi}\left(F^{u\sigma}F_{\nu\sigma} - \tfrac{1}{4}\delta^\mu{}_\nu F^{\tau\sigma}F_{\tau\sigma}\right)_{;\mu} =$$
$$\tfrac{\kappa}{4\pi}\left(F_{\nu\mu}F^{\sigma\mu}{}_{;\sigma} + *F_{\nu\mu}*F^{\sigma\mu}{}_{;\sigma}\right) = \tfrac{\kappa}{4\pi}\left[\tfrac{1}{2}F^{\mu\sigma}\left(F_{\mu\nu;\sigma} + F_{\nu\sigma;\mu} + F_{\sigma\mu;\nu}\right) + F_{\nu\mu}F^{\sigma\mu}{}_{;\sigma}\right] = 0 \tag{4.5}$$

Now the question arises, how does one obtain solutions to the Einstein equation, for the electromagnetic field tensor (4.2)? We know, of course, that we obtain solutions using equation (4.2) itself, which is a second order differential equation for the spacetime metric $g_{\mu\nu}$. But, equation (4.5), which is a *third order* differential equation in $g_{\mu\nu}$, must also be understood as part of the solution. That is, whereas equation (4.2) imposes second order differential conditions on $g_{\mu\nu}$ once the Maxwell tensor has been chosen for examination, equation (4.5) imposes *third order* differential conditions on the spacetime metric, which are also based on the choice of the Maxwell tensor for examination. Now, there are several parts to equation (4.5) that need to be considered. $\left(R^{\mu}{}_{\nu} - \tfrac{1}{2}\delta^{\mu}{}_{\nu}R\right)_{;\mu} = \tfrac{\kappa}{4\pi}\left(F^{u\sigma}F_{\nu\sigma} - \tfrac{1}{4}\delta^{\mu}{}_{\nu}F^{\tau\sigma}F_{\tau\sigma}\right)_{;\mu}$ is merely a trivial, differentiated restatement of (4.2), and so adds no further constraints to the solution. The Bianchi identity $\left(R^{\mu}{}_{\nu} - \tfrac{1}{2}\delta^{\mu}{}_{\nu}R\right)_{;\mu} = 0$ is a mathematical identity which also adds no constraints beyond what has long been known.

However, the relations in the bottom line of (4.5) are non-trivial, and they implicitly incorporate the identity (2.2), $\tfrac{1}{2}F^{\mu\sigma}\left(F_{\mu\nu;\sigma} + F_{\nu\sigma;\mu} + F_{\sigma\mu;\nu}\right) - *F_{\nu\mu} * F^{\sigma\mu}{}_{;\sigma} = 0$. If we take the next step to substitute the electric and magnetic monopole currents and the third rank $P_{\tau\sigma\nu}$ from (1.7), (1.10) and (1.11), we find that:

$$\kappa \cdot \kappa_{\nu\,Maxwell} = \left(R^{\mu}{}_{\nu} - \tfrac{1}{2}\delta^{\mu}{}_{\nu}R\right)_{;\mu} = \kappa\left(F_{\nu\mu}J^{\mu} + *F_{\nu\mu}P^{\mu}\right) = \kappa\left(\tfrac{1}{2}F^{\mu\sigma}P_{\mu\nu\sigma} + F_{\nu\mu}J^{\mu}\right) = 0. \qquad (4.6)$$

This imposes a *non-trivial* set of *third order differential constraints* on the metric tensor $g_{\mu\nu}$, based on the electric charge current $J^{\mu}$, the magnetic charge current $P^{\mu}$, the field strength tensor $F_{\nu\mu}$, and the third rank antisymmetric tensor $P_{\mu\nu\sigma}$.[*]

In this light, we come to view General Relativity as entailing not only solutions to a second order differential equation driven by an energy tensor of matter $T^{\mu}{}_{\nu}$, but also, as entailing solutions to a *third order differential equation* driven by the $\kappa_{\nu}$ associated with $T^{\mu}{}_{\nu}$, that is, as entailing the *pair of equations*:

$$\begin{cases} R^{\mu}{}_{\nu} - \tfrac{1}{2}\delta^{\mu}{}_{\nu}R = -\kappa T^{\mu}{}_{\nu} \\ \left(R^{\mu}{}_{\nu} - \tfrac{1}{2}\delta^{\mu}{}_{\nu}R\right)_{;\mu} = \kappa \cdot \kappa_{\nu}(=0) \end{cases}. \qquad (4.7)$$

For a particular *choice* of energy tensor $T^{\mu}{}_{\nu}$ (of which $T^{\mu}{}_{\nu\,Maxwell}$ in (4.2) is but one example), and for the particular $\kappa_{\nu}$ which that choice of $T^{\mu}{}_{\nu}$ concurrently implies via the identities (2.2) through (2.5), (of which $\kappa_{\nu\,Maxwell}$ in (4.6) is but one example), we impose additional constraints on the solution, at third order, using the $\kappa_{\nu}$. The "zero" to which $\kappa_{\nu}$ is set for any given energy tensor, is the zero of the Bianchi identity, and it drives solutions toward the equilibrium state defined by $\kappa_{\nu} = 0$ for the particular chosen $\kappa_{\nu}$ and its associated $T^{\mu}{}_{\nu}$. The zeros of identities (2.2) through (2.5) – as opposed to the zero of the Bianchi identity – are used to deduce a non-trivial $\kappa_{\nu}$ for the particular energy tensor of choice, so the appropriate third order constraints may then be imposed.

To deepen this perspective, let us pursue a similar line of analysis with equation (2.14) for the trace matter tensor $\mathfrak{J}^{\mu}{}_{\nu}$. Similar to (4.1), we remove the Bianchi identity and introduce $\tau^{\mu}{}_{\nu}$ in place of $t(A)^{\mu}{}_{\nu}$ to write:

---

[*] In [5] in a footnote on page 28, this author notes that these third rank antisymmetric sources appear to contain exactly three sources of "charge," and so may upon careful examination and development come to be understood as describing baryons currents, e.g., $J^{\mu} = \tfrac{1}{3!}\varepsilon^{\alpha\tau\gamma\mu}J_{\alpha\tau\gamma}$ (see 1.12) in covariant fashion, just as $J^{\mu} = \bar{\psi}\gamma^{\mu}\psi$ covariantly describes Fermions.

$$-\kappa T^{\mu}{}_{\nu;\mu} = -\kappa\left(\mathfrak{I}^{\mu}{}_{\nu} + \tau^{\mu}{}_{\nu}\right)_{;\mu} = -\kappa\mathfrak{I}^{\mu}{}_{\nu;\mu} - \kappa\cdot\kappa_{\nu\,trace\,matter}$$
$$= -\tfrac{\kappa}{4\pi}\left[-\tfrac{1}{2}*F^{\mu\sigma}\left(F_{\mu\nu;\sigma} + F_{\nu\sigma;\mu} + F_{\sigma\mu;\nu}\right) - F_{\nu\mu}*F^{\sigma\mu}{}_{;\sigma}\right] \qquad (4.8)$$
$$= -\tfrac{\kappa}{4\pi}\left[-\left(*F^{\mu\sigma}F_{\nu\sigma} - \tfrac{1}{2}\delta^{\mu}{}_{\nu}*F^{\sigma\tau}F_{\sigma\tau}\right)_{;\mu} - \tfrac{1}{2}F^{\sigma\tau}*F_{\sigma\tau;\nu}\right] = 0$$

Now, we identify the $\mathfrak{I}^{\mu}{}_{\nu}$ of (2.15) as the material tensor of interest, and so we associate this with $R^{\mu}{}_{\nu} - \tfrac{1}{2}\delta^{\mu}{}_{\nu}R$ using:

$$R^{\mu}{}_{\nu} - \tfrac{1}{2}\delta^{\mu}{}_{\nu}R = -\kappa\mathfrak{I}^{\mu}{}_{\nu} = \tfrac{\kappa}{4\pi}\left(*F^{\mu\sigma}F_{\nu\sigma} - \tfrac{1}{2}\delta^{\mu}{}_{\nu}*F^{\sigma\tau}F_{\sigma\tau}\right) \qquad (4.9)$$

which from identity (2.5), $-\tfrac{1}{2}*F^{\mu\sigma}\left(F_{\mu\nu;\sigma} + F_{\nu\sigma;\mu} + F_{\sigma\mu;\nu}\right) - F_{\nu\mu}*F^{\sigma\mu}{}_{;\sigma}$ concurrently yields:

$$\kappa_{\nu\,Trace\,Matter} = -\tfrac{1}{8\pi}F^{\sigma\tau}*F_{\sigma\tau;\nu}. \qquad (4.10)$$

Next, we take contracted covariant derivative of each side of (4.9), to yield:

$$-\kappa\mathfrak{I}^{\mu}{}_{\nu;\mu} = \tfrac{\kappa}{4\pi}\left(*F^{\mu\sigma}F_{\nu\sigma} - \tfrac{1}{2}\delta^{\mu}{}_{\nu}*F^{\sigma\tau}F_{\sigma\tau}\right)_{;\mu} = \left(R^{\mu}{}_{\nu} - \tfrac{1}{2}\delta^{\mu}{}_{\nu}R\right)_{;\mu} = 0, \qquad (4.11)$$

Then, we combine (4.8) and (4.11), using the $1 \Leftrightarrow *$ symmetry of (2.33), to yield:

$$\left(R^{\mu}{}_{\nu} - \tfrac{1}{2}\delta^{\mu}{}_{\nu}R\right)_{;\mu} = -\kappa\mathfrak{I}^{\mu}{}_{\nu;\mu} = \kappa\cdot\kappa_{\nu\,Trace\,Matter} = \tfrac{\kappa}{4\pi}\left(*F^{\mu\sigma}F_{\nu\sigma} - \tfrac{1}{2}\delta^{\mu}{}_{\nu}*F^{\sigma\tau}F_{\sigma\tau}\right)_{;\mu},$$
$$= -\tfrac{\kappa}{8\pi}F^{\sigma\tau}*F_{\sigma\tau;\nu} = -\tfrac{\kappa}{4\pi}\left[*F^{\mu\sigma}F_{\mu\nu;\sigma} - \tfrac{1}{2}*F^{\mu\sigma}\left(F_{\mu\nu;\sigma} + F_{\nu\sigma;\mu} + F_{\sigma\mu;\nu}\right)\right] = 0 \qquad (4.12)$$

Finally, we employ (1.10) and (2.34) in the above to yield ($\left(F^{\sigma\tau}*F_{\sigma\tau}\right)_{;\nu} = \left(F^{\sigma\tau}*F_{\sigma\tau}\right)_{,\nu}$, though in general we will retain covariant derivatives of this term):

$$\kappa\cdot\kappa_{\nu\,Trace\,Matter} = \left(R^{\mu}{}_{\nu} - \tfrac{1}{2}\delta^{\mu}{}_{\nu}R\right)_{;\mu} = -\tfrac{\kappa}{16\pi}\left(F^{\sigma\tau}*F_{\sigma\tau}\right)_{;\nu} = \tfrac{1}{2}\kappa*F^{\mu\sigma}P_{\mu\nu\sigma} - \tfrac{\kappa}{4\pi}*F^{\mu\sigma}F_{\mu\nu;\sigma} = 0, \qquad (4.13)$$

To obtain solutions to Einstein's equation, we employ (4.9) and (4.13) in the pair of second and third order Einstein equations (4.7).

From here, we deduce the $\kappa_{\nu}$ for the remaining energy tensors (3.17) to (3.20). Recall from (3.8) that $T^{\mu\nu}{}_{Euler} = T^{\mu\nu}{}_{Maxwell} + \mathfrak{I}^{\mu\nu}$, thus $T^{\mu\nu}{}_{Euler;\mu} = T^{\mu\nu}{}_{Maxwell;\mu} + \mathfrak{I}^{\mu\nu}{}_{;\mu}$. Further, recall from (4.5), $T^{\mu}{}_{\nu\,Maxwell;\mu} = -\kappa_{\nu\,Maxwell}$, and from (4.12), $\mathfrak{I}^{\mu}{}_{\nu;\mu} = -\kappa_{\nu\,Trace\,Matter}$. Further, from (4.1) and (4.8), as a general rule, for any $T^{\mu}{}_{\nu}$ energy tensor, we must set $T^{\mu}{}_{\nu\,Total;\mu} = T^{\mu}{}_{\nu;\mu} + \kappa_{\nu} = 0$, hence $T^{\mu}{}_{\nu;\mu} = -\kappa_{\nu}$. Thus, using (4.6) and (4.13):

$$\kappa_{\nu\,Euler} = \kappa_{\nu\,Maxwell} + \kappa_{\nu\,Trace\,Matter} = F_{\nu\mu}J^{\mu} + *F_{\nu\mu}P^{\mu} - \tfrac{1}{16\pi}\left(F^{\sigma\tau}*F_{\sigma\tau}\right)_{;\nu}$$
$$= \tfrac{1}{2}F^{\mu\sigma}P_{\mu\nu\sigma} + \tfrac{1}{2}*F^{\mu\sigma}P_{\mu\nu\sigma} + F_{\nu\mu}J^{\mu} - \tfrac{1}{4\pi}*F^{\mu\sigma}F_{\mu\nu;\sigma} = 0 \qquad (4.14)$$

So, if the material system under consideration is an adiabatic fluid described by the Euler tensor, and by setting $R^\mu{}_\nu - \frac{1}{2}\delta^\mu{}_\nu R = -\kappa T^\mu{}_{\nu\,Euler}$ we define $T^\mu{}_{\nu\,Euler}$ as the equilibrium state, then $\kappa_{\nu\,Euler}$ will be used in the third order equation (4.7) when the Euler tensor $T^\mu{}_{\nu\,Euler}$ of (3.20) is used in the second order equation (4.7). From the above, we observe the simple, helpful rule that, in general, *the $\kappa_\nu$ will combine according to the same additive rules as their associated energy tensors $T^{\mu\nu}$*. Above, $T^{\mu\nu}{}_{Euler} = T^{\mu\nu}{}_{Maxwell} + \Im^{\mu\nu}$ and thus $\kappa_{\nu\,Euler} = \kappa_{\nu\,Maxwell} + \kappa_{\nu\,Trace\,Matter}$, for example. This additive rule will be helpful in some of the calculations to follow.

Next, recalling (3.25), an equation analogous to (4.1) can be used to derive, (see also (4.14) for the connection to the second line):

$$\kappa_{\nu\,Flux} = \kappa_{\nu\,Matter\,Flux} + \kappa_{\nu\,Pressure\,Flux} = F_{\nu\mu}J^\mu + {}^*F_{\nu\mu}P^\mu + \tfrac{1}{16\pi}\left(F^{\sigma\tau}{}^*F_{\sigma\tau}\right)_{;\nu}$$
$$= \tfrac{1}{2}F^{\mu\sigma}P_{\mu\nu\sigma} + \tfrac{1}{2}{}^*F^{\mu\sigma}P_{\mu\nu\sigma} + F_{\nu\mu}J^\mu - \tfrac{1}{4\pi}{}^*F^{\mu\sigma}F_{\mu\nu;\sigma} + \tfrac{1}{8\pi}\left(F^{\sigma\tau}{}^*F_{\sigma\tau}\right)_{;\nu} = 0 \quad (4.15)$$

Next, comparing (3.18) and (1.5), we note that that $T^{\mu\nu}{}_{Pressure\,Flux} = -\tfrac{1}{2}T^{\mu\nu}{}_{Maxwell}$. Therefore, from (4.6), and using and the additive rule for $T^{\mu\nu}$ and $\kappa_\nu$ developed above, we get:

$$\kappa_{\nu\,Pressure\,Flux} = -\tfrac{1}{2}\kappa_{\nu\,Maxwell} = -\tfrac{1}{2}\left(F_{\nu\mu}J^\mu + {}^*F_{\nu\mu}P^\mu\right) = -\left(\tfrac{1}{4}F^{\mu\sigma}P_{\mu\nu\sigma} + \tfrac{1}{2}F_{\nu\mu}J^\mu\right) = 0. \quad (4.16)$$

Then, using (4.15) and (4.16), we get:

$$\kappa_{\nu\,Matter\,Flux} = \kappa_{\nu\,Flux} - \kappa_{\nu\,Pressure\,Flux} = \tfrac{3}{2}\left(F_{\nu\mu}J^\mu + {}^*F_{\nu\mu}P^\mu\right) + \tfrac{1}{16\pi}\left(F^{\sigma\tau}{}^*F_{\sigma\tau}\right)_{;\nu}$$
$$= \tfrac{3}{4}F^{\mu\sigma}P_{\mu\nu\sigma} + \tfrac{1}{2}{}^*F^{\mu\sigma}P_{\mu\nu\sigma} + \tfrac{3}{2}F_{\nu\mu}J^\mu - \tfrac{1}{4\pi}{}^*F^{\mu\sigma}F_{\mu\nu;\sigma} + \tfrac{1}{8\pi}\left(F^{\sigma\tau}{}^*F_{\sigma\tau}\right)_{;\nu} = 0 \quad (4.17)$$

Next, we note that $T^{\mu\nu}{}_{Stationary\,Pressure} = T^{\mu\nu}{}_{Euler} - T^{\mu\nu}{}_{Flux}$. This contains no sources of either first or third rank, and is simply:

$$\kappa_{\nu\,Stationary\,Pressure} = \kappa_{\nu\,Euler} - \kappa_{\nu\,Flux} = -\tfrac{1}{8\pi}\left(F^{\sigma\tau}{}^*F_{\sigma\tau}\right)_{;\nu} = 0. \quad (4.18)$$

Finally, returning to (2.11) where we derived $T^\mu{}_{\nu\,Vacuum} = 0$, we must now realize, in light of the above, that this is also a non-trivial solution. For, together with:

$$T^\mu{}_{\nu\,Vacuum} = 0, \quad (4.19)$$

we also have:

$$\kappa_{\nu\,Vacuum} = F_{\nu\mu}J^\mu + {}^*F_{\nu\mu}P^\mu = \tfrac{1}{2}F^{\mu\sigma}P_{\mu\nu\sigma} + F_{\nu\mu}J^\mu = 0 (= \kappa_{\nu\,Maxwell}). \quad (4.20)$$

This establishes non-trivial third order constraints (4.7) on the metric, *even when one is obtaining vacuum solutions to the Einstein equation.*

In fact, (4.19) and (4.20) are another example of the way a particular choice of $T^\mu{}_{\nu\,Matter}$ set to $R^\mu{}_\nu - \tfrac{1}{2}\delta^\mu{}_\nu R$ determines the equilibrium point of Einstein's equations. Let us postulate $R^\mu{}_\nu - \frac{1}{2}\delta^\mu{}_\nu R = T^\mu{}_{\nu\,Matter} = T^\mu{}_{\nu\,Vacuum} = 0$ throughout a closed system of energy and therefore, that (4.20) above also applies. Many known solutions in fact make this choice. We also start by postulating $J^\mu = P^\mu = 0$. Now, suppose we disturb this system by introducing a source of electric charge, $J^\mu \neq 0$. Immediately, $\kappa_{\nu\,Vacuum} \cong 0$, but $\kappa_{\nu\,Vacuum} \neq 0$. That is, the gravitational energy exchange is approximately, but no longer exactly equal to zero. In other words, $0 \cong R^\mu{}_\nu - \frac{1}{2}\delta^\mu{}_\nu R \cong T^\mu{}_{\nu\,Vacuum}$, but more to the point:

$$R^\mu{}_\nu - \tfrac{1}{2}\delta^\mu{}_\nu R = T^\mu{}_{\nu\,Matter} = T^\mu{}_{\nu\,Vacuum} + \delta T^\mu{}_\nu = 0 + \delta T^\mu{}_\nu, \tag{4.21}$$

where $\delta T^\mu{}_\nu$ represents a perturbation from the vacuum caused by introducing $J^\mu \neq 0$. Now, the energy tensor is not longer identically $T^\mu{}_{\nu\,Vacuum}$, but rather is a composition of both $T^\mu{}_{\nu\,Vacuum}$ and $\delta T^\mu{}_\nu$. Because we have "disturbed" what we postulated to be a closed system of energy by adding a charge $J^\mu \neq 0$, we have now created a new equilibrium point where the energy tensor is no longer quite $T^\mu{}_{\nu\,Vacuum}$, but instead also includes $\delta T^\mu{}_\nu$, because of the hypothesis that a charge was introduced. Thus, we see that the earlier discussion about equilibrium applies to *closed* systems of energy. The moment we violate the closed system and introduce something new (a charge, a fluid, a rock), then we have implicitly changed the equilibrium point of the system, causing both a $\delta T^\mu{}_\nu$ correction to $T^\mu{}_{\nu\,Matter}$, as well as a $\delta\kappa_\nu$ correction to the gravitational field.

More generally, this means that if we perturb an energy tensor by $T^\mu{}_\nu \rightarrow T^\mu{}_\nu + \delta T^\mu{}_\nu$, i.e., $\kappa_\nu \rightarrow \kappa_\nu + \delta\kappa_\nu$, where the energy of $\delta T^\mu{}_\nu$ is in a *different form* than the energy of $T^\mu{}_\nu$, and if the equilibrium point is defined as $R^\mu{}_\nu - \tfrac{1}{2}\delta^\mu{}_\nu R \equiv T^\mu{}_\nu$, $\kappa_\nu = 0$, then the natural geodesic tendency of this system to seek an equilibrium at $R^\mu{}_\nu - \tfrac{1}{2}\delta^\mu{}_\nu R \equiv T^\mu{}_\nu$, $\kappa_\nu = 0$ will cause the $\delta T^\mu{}_\nu$ form of energy to convert to the $T^\mu{}_\nu$ form of energy through exchanges with the gravitational field. Again, energy conversion is seen to be a *gravitational* phenomenon.

The "=0" for the various $\kappa_\nu$ above is a reminder that each of these is connected with $\left(R^\mu{}_\nu - \tfrac{1}{2}\delta^\mu{}_\nu R\right)_{;\mu} = \kappa \cdot \kappa_\nu (=0)$ in equation (4.7), to impose third-order differential constraints on the spacetime metric, at the same time as their associated energy tensors are used to impose second-order differential constraints via the Einstein equation. Again, however, once a closed energy system is disturbed, then the definition of which particular $T^\mu{}_{\nu\,Matter}$ is set to $R^\mu{}_\nu - \tfrac{1}{2}\delta^\mu{}_\nu R$ and which $\kappa_\nu$ is set to $\left(R^\mu{}_\nu - \tfrac{1}{2}\delta^\mu{}_\nu R\right)_{;\mu}$ is altered. This explains the "freedom" we have to choose from a variety of energy tensors, since the choice of an energy tensor is based on what best describes the material composition of the particular, closed system of energy under consideration. When a change in material composition introduced from outside (as opposed to natural, geodesic changes within a closed system seeking equilibrium), there is also introduced a change in energy tensor as well. What that system will then do as a dynamical matter, depends on which energy tensor is associated with $R^\mu{}_\nu - \tfrac{1}{2}\delta^\mu{}_\nu R$, that is, on the definition of the equilibrium $\kappa_\nu = 0$ whereby no energy is exchanged between matter and the gravitational field.

Now, we turn to discuss how solutions to the Einstein equation based on the energy tensors $T^{\mu}{}_{\nu}$ elaborated in section 3 and their associated $\kappa_{\nu}$ just derived, may lead a quantum theory of gravitation and matter.

## 5. Conclusion: The Road to Quantum Gravitation

At this point, armed with the energy tensors $T^{\mu\nu}$ developed in the section 3 and the $\kappa_{\nu}$ from section 4, we consider what sorts of solutions to the paired, second and third order Einstein equations (4.7), $-\kappa T_{\mu\nu} = R_{\mu\nu} - \frac{1}{2} g_{\mu\nu} R$ and $\left(R^{\mu}{}_{\nu} - \frac{1}{2}\delta^{\mu}{}_{\nu} R\right)_{;\mu} = \kappa \cdot \kappa_{\nu}$, might be obtained.

First, we note the very important fact that the results derived thus far do not at all depend upon whether the field strength tensors are Abelian as in (1.4), or non-Abelian as in (1.3). The identity (2.1) is based on the antisymmetric character of tensors $A^{\sigma\tau}$ and $B^{\tau\sigma}$, together with the definition $*A^{\sigma\tau} \equiv \frac{1}{2!}\varepsilon^{\delta\gamma\sigma\tau}A_{\delta\gamma}$, and thus applies to *any* antisymmetric second rank tensor, Abelian or not. Thus, as noted after (2.1), $\frac{1}{2}A^{i\sigma\tau}\left(B_{i\tau\sigma;\nu} + B_{i\sigma\nu;\tau} + B_{i\nu\tau;\sigma}\right) - *A_{i\nu\sigma}*B^{i\tau\sigma}{}_{;\tau} = 0$ will also hold true for *any* dual tensor defined as $*A_i{}^{\sigma\tau} \equiv \frac{1}{2!}\varepsilon^{\delta\gamma\sigma\tau}A_{i\delta\gamma}$. Because of this, every single field tensor in sections 2, 3, and 4 can be generalized to $F^{\mu\nu} \Rightarrow F_i{}^{\mu\nu}$, $*F^{\mu\nu} \Rightarrow *F_i{}^{\mu\nu}$ every current can be generalized to $J^{\mu} \Rightarrow J_i{}^{\mu}$, $P^{\mu} \Rightarrow P_i{}^{\mu}$, and the third rank antisymmetric tensors can be generalized to $J_{\mu\nu\tau} \Rightarrow J_{i\mu\nu\tau}$, $P_{\mu\nu\tau} \Rightarrow P_{i\mu\nu\tau}$. Thus, we now summarize the results this far, generalized to non-Abelian gauge theory, as follows. From (3.2) and (4.6):

$$\begin{cases} T^{\mu\nu}{}_{Maxwell} = -\frac{1}{4\pi}\left[F^{i\mu\sigma}F_i{}^{\nu}{}_{\sigma} - \frac{1}{4}g^{\mu\nu}F^{i\tau\sigma}F_{i\tau\sigma}\right] \\ \kappa_{\nu\,Maxwell} = F_{i\nu\mu}J^{i\mu} + *F_{i\nu\mu}P^{i\mu} = F_{i\nu\mu}J^{i\mu} + \frac{1}{2}F^{i\mu\sigma}P_{i\mu\nu\sigma} = 0 \end{cases} \quad (5.1)$$

From (3.3) and (4.13):

$$\begin{cases} \mathfrak{I}^{\mu\nu} = -\frac{1}{4\pi}\left(*F^{i\mu\sigma}F_i{}^{\nu}{}_{\sigma} - \frac{1}{2}g^{\mu\nu}*F^{i\tau\sigma}F_{i\sigma\tau}\right) \\ \kappa_{\nu\,Trace\,Matter} = -\frac{1}{16\pi}\left(F^{i\sigma\tau}*F_{i\sigma\tau}\right)_{;\nu} = \frac{1}{2}*F^{i\mu\sigma}P_{i\mu\nu\sigma} - \frac{1}{4\pi}*F^{i\mu\sigma}F_{i\mu\nu;\sigma} = 0 \end{cases} \quad (5.2)$$

From (3.20) and (4.14):

$$\begin{cases} T^{\mu\nu}{}_{Euler} = -\frac{1}{4\pi}\left[F^{i\mu\sigma}F_i{}^{\nu}{}_{\sigma} + *F^{i\mu\sigma}F_i{}^{\nu}{}_{\sigma} - \frac{1}{4}g^{\mu\nu}F^{i\tau\sigma}F_{i\tau\sigma} - \frac{1}{2}g^{\mu\nu}*F^{i\sigma\tau}F_{i\sigma\tau}\right] \\ \kappa_{\nu\,Euler} = F_{i\nu\mu}J^{i\mu} + *F_{i\nu\mu}P^{i\mu} - \frac{1}{16\pi}\left(F^{i\sigma\tau}*F_{i\sigma\tau}\right)_{;\nu} \\ = \frac{1}{2}F^{i\mu\sigma}P_{i\mu\nu\sigma} + \frac{1}{2}*F^{i\mu\sigma}P_{i\mu\nu\sigma} + F_{i\nu\mu}J^{i\mu} - \frac{1}{4\pi}*F^{i\mu\sigma}F_{i\mu\nu;\sigma} = 0 \end{cases} \quad (5.3)$$

From (3.25) and (4.15):

$$\begin{cases} T^{\mu}{}_{\nu\,Flux} = -\frac{1}{4\pi} F^{i\mu\sigma} F_{i\nu\sigma} \\ \kappa_{\nu\,Flux} = F_{i\nu\mu} J^{i\mu} + {}^*F_{i\nu\mu} P^{i\mu} + \frac{1}{16\pi}\left(F^{i\sigma\tau} * F_{i\sigma\tau}\right)_{;\nu} \\ = \frac{1}{2} F^{i\mu\sigma} P_{i\mu\nu\sigma} + \frac{1}{2} * F^{i\mu\sigma} P_{i\mu\nu\sigma} + F_{i\nu\mu} J^{i\mu} - \frac{1}{4\pi} * F^{i\mu\sigma} F_{i\mu\nu;\sigma} + \frac{1}{8\pi}\left(F^{i\sigma\tau} * F_{i\sigma\tau}\right)_{;\nu} = 0 \end{cases} \quad (5.4)$$

From (3.21) and (4.17):

$$\begin{cases} T^{\mu}{}_{\nu\,Matter\,Flux} = -\frac{1}{4\pi}\left[\frac{5}{4} F^{i\mu\sigma} F_i{}^{\nu}{}_{\sigma} + \frac{1}{4} * F^{i\mu\sigma} F_i{}^{\nu}{}_{\sigma}\right] \\ \kappa_{\nu\,Matter\,Flux} = \frac{3}{2}\left(F_{i\nu\mu} J^{i\mu} + {}^*F_{i\nu\mu} P^{i\mu}\right) + \frac{1}{16\pi}\left(F^{i\sigma\tau} * F_{i\sigma\tau}\right)_{;\nu} \\ = \frac{3}{4} F^{i\mu\sigma} P_{i\mu\nu\sigma} + \frac{1}{2} * F^{i\mu\sigma} P_{i\mu\nu\sigma} + \frac{3}{2} F_{i\nu\mu} J^{i\mu} - \frac{1}{4\pi} * F^{i\mu\sigma} F_{i\mu\nu;\sigma} + \frac{1}{8\pi}\left(F^{i\sigma\tau} * F_{i\sigma\tau}\right)_{;\nu} = 0 \end{cases} \quad (5.5)$$

From (3.22) and (4.16):

$$\begin{cases} T^{\mu\nu}{}_{Pressure\,Flux} = \frac{1}{4\pi}\left[\frac{1}{4} F^{i\mu\sigma} F_i{}^{\nu}{}_{\sigma} + \frac{1}{4} * F^{i\mu\sigma} F_i{}^{\nu}{}_{\sigma}\right] \\ \kappa_{\nu\,Pressure\,Flux} = -\frac{1}{2}\left(F_{i\nu\mu} J^{i\mu} + {}^*F_{i\nu\mu} P^{i\mu}\right) = -\left(\frac{1}{4} F^{i\mu\sigma} P_{i\mu\nu\sigma} + \frac{1}{2} F_{i\nu\mu} J^{i\mu}\right) = 0 \end{cases} \quad (5.6)$$

From (3.19) and (4.18):

$$\begin{cases} T^{\mu\nu}{}_{Stationary\,Pressure} = -\frac{1}{4\pi}\left[* F^{i\mu\sigma} F_i{}^{\nu}{}_{\sigma} - \frac{1}{4} g^{\mu\nu} F^{i\tau\sigma} F_{i\tau\sigma} - \frac{1}{2} g^{\mu\nu} * F^{i\sigma\tau} F_{i\sigma\tau}\right] \\ \kappa_{\nu\,Stationary\,Pressure} = -\frac{1}{8\pi}\left(F^{i\sigma\tau} * F_{i\sigma\tau}\right)_{;\nu} = 0 \end{cases} \quad (5.7)$$

Finally, from (4.19) and (4.20):

$$\begin{cases} T^{\mu}{}_{\nu\,Vacuum} = 0 \\ \kappa_{\nu\,Vacuum} = F_{i\nu\mu} J^{i\mu} + {}^*F_{i\nu\mu} P^{i\mu} = \frac{1}{2} F^{i\mu\sigma} P_{i\mu\nu\sigma} + F_{i\nu\mu} J^{i\mu} = 0 \end{cases} \quad (5.8)$$

All of the above do not by any means rule out the development and use of additional energy tensors and associated gravitational components for other material phenomena.

Each of the above is used pairwise, in connection to the Einstein equations of (4.7):

$$\begin{cases} -\kappa T^{\mu}{}_{\nu} = R^{\mu}{}_{\nu} - \frac{1}{2} \delta^{\mu}{}_{\nu} R = \\ \kappa \cdot \kappa_{\nu} = \left(R^{\mu}{}_{\nu} - \frac{1}{2} \delta^{\mu}{}_{\nu} R\right)_{;\mu} (= 0) \end{cases} \quad (5.9)$$

to establish second and third order conditions on the metric tensor $g_{\mu\nu}$, based on the material phenomena under consideration.

Now, we come to ultimate question, how do we use this to establish a connection to quantum gravitation? More to the point, how do we use this to establish non-linearity for all of the known quantum mechanical interactions, based on gravitational theory? And, how in the process do we "geometrize" quantum mechanics and its wavefunctions, or, conversely, quantize gravitational geometry to yield metric wavefunctions which have discrete sets of probability distributions based on underlying sets of quantum numbers?

Although obtaining complete, exact quantum solutions for $g_{\mu\nu}$ may be difficult in practice, in theory the recipe can be defined in a straightforward manner:

First, for all vector bosons $A^{i\mu}$ of given mass M in the field of a current $J^{i\nu}$, we solve the relativistic Schrödinger (Klein-Gordon) equation:

$$J^{i\nu} = \tfrac{1}{4\pi}\left(A^{i\nu;\mu}{}_{;\mu} + M^2 A^{i\nu}\right) \tag{5.10}$$

and thereby represent $A^{i\mu}$ as a vector boson wavefunction. Alternatively, using (1.4) and (1.7), it may be suitable to write the wave equation with $A^{i\mu}{}_{;\mu} = 0$ as:

$$J^{i\nu} = \tfrac{1}{4\pi} F^{i\mu\nu}{}_{;\mu} = \tfrac{1}{4\pi}\left(A^{i\nu;\mu} - A^{i\mu;\nu} - gf^{ijk} A_j{}^\mu A_k{}^\nu\right)_{;\mu} = \tfrac{1}{4\pi}\left(A^{i\nu;\mu}{}_{;\mu} - gf^{ijk} A_j{}^\mu A_k{}^\nu{}_{;\mu}\right). \tag{5.11}$$

Comparing (5.10) and (5.11) above, it is worth noting that $M^2 A^{i\nu} \sim -gf^{ijk} A_j{}^\mu A_k{}^\nu{}_{;\mu}$, i.e., that $-gf^{ijk} A_j{}^\mu A_k{}^\nu{}_{;\mu}$ plays a role in (5.11) analogous to that of $M^2 A^{i\nu}$ in (5.10), which leads to the suspicion (worthy of investigation in its own right) that this term $-gf^{ijk} A_j{}^\mu A_k{}^\nu{}_{;\mu}$ is somehow involved in the symmetry breaking mechanism that leads to vector boson masses.

Next, for all Fermions $\psi$, we solve the Dirac equation:

$$\left(i\gamma^\mu\left(\partial_\mu + g\lambda^i A_\mu\right) - m\right)\psi = 0 \tag{5.12}$$

together with the adjoint equation for $\overline{\psi}$, thereby obtaining the fermion wavefunctions and probability distributions / densities which, of course, have discrete configurations based on the quantum numbers of the fermion.

Then, we take the $A^{i\mu}$ wavefunctions from (5.10) or (5.11) and define the fields according to either (1.3):

$$F_{\mu\nu} = A_{\nu;\mu} - A_{\mu;\nu} \tag{5.13}$$

or (1.4):

$$F_{i\mu\nu} = A_{i\nu;\mu} - A_{i\mu;\nu} - gf_{ijk} A^j{}_\mu A^k{}_\nu, \tag{5.14}$$

depending on whether the interaction is Abelian or non-Abelian, and, if non-Abelian, using the appropriate group structure constants $f_{ijk}$ and charge $g$. Now, the fields $F_{\mu\nu}$, $F_{i\mu\nu}$ themselves are quantum wavefunctions.

Now we turn to the Fermion wavefunctions, and form a current four-vector

$$J^{i\mu} = g\overline{\psi}\gamma^\mu \lambda^i \psi \tag{5.15}$$

from the Dirac equation solutions that we have obtained for $\psi$, $\overline{\psi}$.

Next, we select the energy tensor $T^{\mu\nu}$ and associated $\kappa_\nu$ from (5.1) through (5.8) corresponding to the particular material phenomenon we wish to examine, and insert the quantized fields from (5.13) through (5.15). This gives us a quantum wavefunction representation of $T^{\mu\nu}$, which is used for the second-order equation in (5.9) and of $\kappa_\nu$ for the third-order equation. This quantum wavefunction representations of $T^{\mu\nu}$ and $\kappa_\nu$ will then implicitly include the quantum characteristics and probability densities of the Fermions, and so the quantum mechanical characteristics of the Fermions will have their effects most directly on the third order derivatives of the spacetime metric.[*] This is why the development of the third-order equation $\kappa \cdot \kappa_\nu = \left(R^\mu{}_\nu - \tfrac{1}{2}\delta^\mu{}_\nu R\right)_{;\mu}$ and of the various $\kappa_\nu$ in the preceding section, is of such importance. Of course, for example, the solution to the wave equation (5.10) for a photon is $A^\mu = -(1/q^2)J^\mu = -(1/q^2)g\overline{\psi}\gamma^\mu\psi$, and so the currents may also enter via (5.13) or (5.14) and thus via $T^{\mu\nu}$ at second order as well.

Finally, we solve the Einstein equations using both second order quantum wavefunction for $T^{\mu\nu}$, and the third-order quantum wavefunction for $\kappa_\nu$. The metric derived therefrom, will also have the characteristics of a quantum wavefunction, and will reflect the quantum wavefunctions introduced by the $T^{\mu\nu}$ and the $\kappa_\nu$. The $T^{\mu\nu}$ and the $\kappa_\nu$ thus define the "interface" between gravitational theory and quantum mechanics. All of this, emanates from the fact that it is possible to describe a wide variety of material energy tensors $T^{\mu\nu}$ and gravitational energy exchange vectors $\kappa_\nu$ *all in terms of* $F^{i\mu\nu}$ and sources $J_i^\mu$, $P_i^\mu$ which may also all be expressed in terms of $F^{i\mu\nu}$, see (5.1) to (5.8) above. And, it emanates from the fact that these inputs to $T^{\mu\nu}$ and $\kappa_\nu$ are all well-understood as quantum mechanical fields. *This is quantum gravitation.*

From this, we learn several things. First, this means that initially, we will likely learn about quantum gravitation from what we already know about the quantum theories of particles and fields based on the Dirac and Klein-Gordon equations operating on fermion and boson wavefunctions. Second, by injecting the boson and fermion wavefunctions into gravitational theory via the $T^{\mu\nu}$ and $\kappa_\nu$ "interface tensors," we may come to understand quantum mechanics in the context of non-linear field theory. Third, depending on the actual solutions arrived at, that is, depending on the character of the spacetime metric once it becomes quantized, we may be able, from that geometric standpoint, to learn other things about quantum mechanics which are presently out of reach in the absence of a geometrized quantum theory. Fourth, because the quantum theory of gravitation and matter as developed here accommodates non-Abelian gauge theories, it will be possible to incorporate weak and strong interactions, together with electroweak interactions, completely into the context of gravitational theory. About the only mystery that seems still out of focus in this light is the question of how the electroweak and strong interactions are to be unified. [8][**] Fifth, the non-linear dynamics of all of these interactions, and the manner by which a system of energy progresses toward stable, equilibrium configurations and converts energy from one form to another and accommodates disruptions in its balance, will be a natural outgrowth of geometrizing quantum mechanics. Sixth, since the dynamics of Einstein's equations are intimately connected with the geodesic equation of motion, this means that when various energy configurations seek equilibrium and convert from one form or energy to another to do so, this progression is ultimately governed by principles of least action and geodesic dynamics in a geometrodynamic context. This is to say, the tendency of a system of energy to dynamically seek equilibrium and convert one form of energy into another in the process of seeking equilibrium is an extension of the geodesic principle to govern the behavior of entire systems of energy, and is at bottom a gravitational phenomenon.

---

[*] If it proves possible to derive baryon wavefunctions from $J_{\mu\nu\tau}, J_{i\mu\nu\tau}$ and $P_{\mu\nu\tau}, P_{i\mu\nu\tau}$, then these too would be used to establish the $\kappa_\nu$.

[**] See, in particular, [8], Section 12.2 for an excellent discussion of how the weak isospin and color symmetries may be connected in relation to lepto-quark unification.

While we shall not attempt here to obtain any quantum solutions for the spacetime metric, will give a rudimentary example of how one would approach obtaining a quantum solution, using the simple "unphysical" example of a static, spinless, spherically-symmetric electron $\phi$, interacting with a massless field of photons $A^\mu$. ([9] takes the same approach in Chapter 4.) If we postulate that we are looking at this single, spinless electron as a closed system of energy which contains only the electron and its electromagnetic fields, and if we neglect the electron's rest mass which has a trace and thus would motivate some use of $\Im^{\mu\nu}$, the energy tensor and gravitational energy vector to be employed are the traceless Maxwell tensor $T^{\mu\nu}{}_{Maxwell}$ and $\kappa_{\nu\,Maxwell}$ from (5.1). Because electromagnetism in an Abelian interaction, we would use (5.1) without the internal symmetry index. We further assume that there are no magnetic charges, that is $P^\sigma = 0$.

Because we have postulated our electron to be spinless and spherically symmetric, one would first obtain *time-independent* and *spherically-symmetric* solutions of the Klein-Gordon equation, in the form:

$$\left(\partial^\mu \partial_\mu + e^2 A^\mu A_\mu + m^2\right)\phi = 0 \tag{5.16}$$

For a real electron, one would of course use the Dirac equation $\left(i\gamma^\mu(\partial_\mu + eA_\mu) - m\right)\psi = 0$ (5.12), and would need to at least relax our postulate to a stationary but not static solution because of the electron's spin and the reversal that would take place in the spin magnetic field under a time reversal. Similarly, we would have to relax our postulate to at best, spherical symmetry about the z-axis (the spin axis), and in higher-energy states with radial nodes, we would have to discard or severely limits all spherical symmetries as well. So, even when we come to look at a real, spin ½ electron (which are an ideal candidate for the first effort to obtain "real" quantum solutions to the Einstein Equations), it is best to start in ground states with no angular momentum other than spin, which provide the highest degree of spherical symmetry and the simplest behavior as a function of time.

Then, in our very simple, unphysical example, we would establish the electric current as:

$$J^\mu = -ie\left(\phi^*(\partial^\mu \phi) - (\partial^\mu \phi^*)\phi\right) \tag{5.17}$$

Similarly, for the photon, we would obtain solutions to the wave equation:

$$\partial^\mu \partial_\mu A^\mu = J^\mu. \tag{5.18}$$

We would then write the field strength tensor as $F_{\mu\nu} = A_{\nu;\mu} - A_{\mu;\nu}$. These would all go into $T^{\mu\nu}{}_{Maxwell}$ and $\kappa_{\nu\,Maxwell}$ in (5.1), without the "i" index. In particular, the paired solution of the Einstein equations would employ:

$$\begin{cases} T^{\mu}{}_{\nu\,Maxwell} = -\frac{1}{4\pi}\left[F^{\mu\sigma}F_{\nu\sigma} - \frac{1}{4}\delta^{\mu}{}_{\nu}F^{\tau\sigma}F_{\tau\sigma}\right] \\ = -\frac{1}{4\pi}\left[(A^{\sigma;\mu} - A^{\mu;\sigma})(A_{\sigma;\nu} - A_{\nu;\sigma}) - \frac{1}{4}\delta^{\mu}{}_{\nu}(A^{\sigma;\tau} - A^{\tau;\sigma})(A_{\sigma;\tau} - A_{\tau;\sigma})\right], \\ \kappa_{\nu\,Maxwell} = F_{\nu\sigma}J^{\sigma} = -ie(A_{\sigma;\nu} - A_{\nu;\sigma})(\phi^*(\partial^{\sigma}\phi) - (\partial^{\sigma}\phi^*)\phi) = 0 \end{cases} \quad (5.19)$$

Because we are postulating a static, spherically symmetric solution, we may use the Schwarzschild solution for a spherically-symmetric, static distribution of energy. For a "real" electron with spin ½, the Kerr solution or a variant thereof which assumes spherical symmetry about the spin (z) axis only, and perhaps is also time dependent, would be more suitable for obtaining ground state solutions. We follow the usual Schwarzschild procedure of defining the metric tensor in terms of unknown parameters N(r) and L(r) according to (see, e.g., [10], section 8.1):

$$g_{\mu\nu} = \begin{pmatrix} e^N & 0 & 0 & 0 \\ 0 & -e^L & 0 & 0 \\ 0 & 0 & -r^2 & 0 \\ 0 & 0 & 0 & -r^2\sin^2\theta \end{pmatrix}, \quad (5.20)$$

with coordinates $x^{\mu} = (t \quad r \quad \theta \quad \phi)$. Then, after calculating Christoffel symbols and the Ricci tensor, using $T^{\mu}{}_{\nu\,Maxwell}$ from (5.19), we would write four simultaneous second order equations:

$$-\kappa T^0{}_{0\,Maxwell} = \frac{\kappa}{4\pi}\left[F^{0\sigma}F_{0\sigma} - \frac{1}{4}F^{\tau\sigma}F_{\tau\sigma}\right] = R^0{}_0 - \frac{1}{2}R = -e^{-L}\left(\frac{L'}{r} - \frac{1}{r^2}\right) - \frac{1}{r^2}.$$
$$= -\frac{1}{4\pi}\left[(A^{\sigma;0} - A^{0;\sigma})(A_{\sigma;0} - A_{0;\sigma}) - \frac{1}{4}(A^{\sigma;\tau} - A^{\tau;\sigma})(A_{\sigma;\tau} - A_{\tau;\sigma})\right] \quad (5.21)$$

$$-\kappa T^1{}_{1\,Maxwell} = \frac{\kappa}{4\pi}\left[F^{1\sigma}F_{1\sigma} - \frac{1}{4}F^{\tau\sigma}F_{\tau\sigma}\right] = R^1{}_1 - \frac{1}{2}R = e^{-L}\left(\frac{N'}{r} + \frac{1}{r^2}\right) - \frac{1}{r^2}.$$
$$= -\frac{1}{4\pi}\left[(A^{\sigma;1} - A^{1;\sigma})(A_{\sigma;1} - A_{1;\sigma}) - \frac{1}{4}(A^{\sigma;\tau} - A^{\tau;\sigma})(A_{\sigma;\tau} - A_{\tau;\sigma})\right] \quad (5.22)$$

$$-\kappa T^2{}_{2\,Maxwell} = \frac{\kappa}{4\pi}\left[F^{2\sigma}F_{2\sigma} - \frac{1}{4}F^{\tau\sigma}F_{\tau\sigma}\right] = R^2{}_2 - \frac{1}{2}R = e^{-L}\left(\frac{N''}{2} - \frac{L'N'}{4} + \frac{N'^2}{4} + \frac{N'-L'}{2r}\right).$$
$$= -\frac{1}{4\pi}\left[(A^{\sigma;2} - A^{2;\sigma})(A_{\sigma;2} - A_{2;\sigma}) - \frac{1}{4}(A^{\sigma;\tau} - A^{\tau;\sigma})(A_{\sigma;\tau} - A_{\tau;\sigma})\right] \quad (5.23)$$

$$-\kappa T^3{}_{3\,Maxwell} = \frac{\kappa}{4\pi}\left[F^{3\sigma}F_{3\sigma} - \frac{1}{4}F^{\tau\sigma}F_{\tau\sigma}\right] = R^3{}_3 - \frac{1}{2}R = e^{-L}\left(\frac{N''}{2} - \frac{L'N'}{4} + \frac{N'^2}{4} + \frac{N'-L'}{2r}\right).$$
$$= -\frac{1}{4\pi}\left[(A^{\sigma;3} - A^{3;\sigma})(A_{\sigma;3} - A_{3;\sigma}) - \frac{1}{4}(A^{\sigma;\tau} - A^{\tau;\sigma})(A_{\sigma;\tau} - A_{\tau;\sigma})\right] \quad (5.24)$$

where $N' = \partial N / \partial r$, $L' = \partial L / \partial r$. Again, the $F_{\mu\nu} = A_{\nu;\mu} - A_{\mu;\nu}$ are based on the $A_\mu$ wavefunctions, which may also be given by $A^\mu = -(1/q^2)J^\mu = ie(1/q^2)(\phi^*(\partial^\mu\phi) - (\partial^\mu\phi^*)\phi)$ (and by $A^\mu = -(1/q^2)J^\mu = -(1/q^2)e\bar{\psi}\gamma^\mu Q\psi$ for a "real" electron), and so bring the electron probability densities into play.

For the third order Einstein equation, because we have postulated spherical symmetry and time independence, $\partial N/\partial t = \partial N/\partial\theta = \partial N/\partial\phi = \partial L/\partial t = \partial L/\partial\theta = \partial L/\partial\phi = 0$, the only non-trivial third order equation is for $\kappa_1$, from (5.22). Using $\kappa_1$ from (5.1), with $P^\mu = 0$, we get:

$$\kappa_{1 Maxwell} = F_{1\mu}J^\mu = \left(R^1{}_1 - \tfrac{1}{2}R\right)_{;1} = \left[-e^{-L}\left(\frac{N''}{r^2} + \frac{L'N'}{r} + \frac{L'}{r^2} + 2\frac{1}{r^3}\right) + 2\frac{1}{r^3}\right]'.$$
$$== -ie(A_{\sigma;1} - A_{1;\sigma})(\phi^*(\partial^\sigma\phi) - (\partial^\sigma\phi^*)\phi)$$
(5.25)

The wavefunction for $J^\mu$ thus sets further constraints on the metric at third order. The $g_{\mu\nu}$ arrived at from these equations (5.21) through (5.25) would then be the metric for an "unphysical" spinless, spherically-symmetric, static electron.

What might be the nature of the spacetime metric once a set of equations like (5.21) through (5.25) are solved? We see here that the second and third derivatives of $g_{\mu\nu}$ are functions of the probability densities of fermions and bosons. At a given point in spaetime, this is related to the probability that the second and third derivatives of $g_{\mu\nu}$ will have a particular value, and / or will be related to the expected value of these derivatives. If an electron is "at that point" there will be one value; if not, there will be another value. The distribution of possible values will therefore be related to the probability of the electron being "at that point" or not. We may surmise therefore, that once a solution for $g_{\mu\nu}$ is obtained, we will have a $g_{\mu\nu}$ wavefunction which, at any given point in spacetime, is *not* a definitive value for $g_{\mu\nu}$, but is rather a probability distribution representing a range of possible values of $g_{\mu\nu}$, and / or a set of expectation values for the $g_{\mu\nu}$. This will, of course, be a function of the quantum numbers of the particles in question. Thus, we will have metrics parameterized by energy n, and angular momentum L, etc. So, for example, while in the classical vacuum Schwarzschild solution, we end up with $g_{00} = 1 - 2GM/r$, $g_{11} = 1/(1 - 2GM/r)$, here we would find that the value of $g_{00}$ at each point in spacetime has a probability density, together with an expectation value. So, the $g_{00} = 0$, $g_{11} = \infty$ singularity at $r = 2GM$ may well turn out to be one of the values in a distributed range of possible values for $g_{00}$, which would cast a somewhat different light on the nature of singularities and event horizons. All this will be best understood once we have been able to obtain definitive, exact solutions for real particles, according the recipe outlined above.

In theory, however, we now have a complete prescription for obtaining quantum solutions to the Einstein equation and for ensuring that these solutions are in complete accord with known principles of quantum mechanics. Clearly, it will be of significant interest, for example, to employ the $F_i^{\mu\nu}$, i = 1, 2, 3, . . . 8 of quantum chromodynamics, to see what sorts of solutions emerge when an SU(3) color triplet of fermions (quarks), via gauge symmetry, drive the gluons $A^i{}_\mu$ and the currents $J^{i\mu}$ and fields $F_i{}^{\mu\nu}$, which in turn drive the "interface" energy tensor $T_{\mu\nu}$ and gravitational components $\kappa_\nu$, which finally drive the metric $g_{\mu\nu}$.

Perhaps the most profound question of all, is what effect, if any, all of this will have on questions of gravitational collapse and singularities, and ultimately, on questions involving the ultimate origin and destiny of the cosmological universe.

## Acknowledgements

The author wishes to recognize and thank Ken S. Tucker for his collaboration and assistance and encouragement during the development of the paper. After reviewing the author's earlier papers [5] and [7], he first pointed out to the author how the gravitational scalar $\sqrt{-g}$ appeared as a natural consequence of duality and that this could indicate a path for connecting general relativity and particle physics. The author then began to take a serious look at how the duality formalism addressed in those earlier papers could be applied directly to gravitational theory, and this paper is the result.

Thanks also to C. Fred Diether for his ongoing encouragement and support of the author's ongoing efforts to acquire a deeper understanding of nature.

## References

[1] Einstein, A., *The Foundation of the General Theory of Relativity*, Annalen der Physik, 49, 1916, in "The Principle of Relativity," Dover (1952), pages 111-164.
[2] Reinich, G.Y., *Electrodynamics in the General Relativity Theory*, Trans. Am. Math. Soc., Vol, **27**, 106-136 (1925).
[3] Wheeler, J. A., *Geometrodynamics*, Academic Press, 225-253 (1962).
[4] Misner, C. W., Thorne, K. S., and Wheeler, J. A., *Gravitation*, W. H. Freeman & Co. (1973)
[5] Yablon, J. R., *Magnetic Monopoles and Duality Symmetry Breaking in Maxwell's Electrodynamics*, hep-ph/0508257 (August 24, 2005).
[6] Einstein, A., *Do Gravitational Fields Play an Essential Part in The Structure of the Elementary Particles of Matter?*, Sitzungsberichte der Preussischen Akad. d. Wissenschafter, 1919, in "The Principle of Relativity," Dover (1952), pages 191-198.
[7] Yablon, J. R., *Magnetic Monopole Interactions, Chiral Symmetries, and the NuTeV Anomaly*, hep-ph/0509223 (September 21, 2005).
[8] Volovik, G. E., *The Universe in a Helium Droplet*, Clarendon Press – Oxford (2003).
[9] Halzen, F., and Martin A. D., *Quarks and Leptons: An Introductory Course in Modern Particle Physics*, J. Wiley & Sons (1984).
[10] Ohanian, H., *Gravitation and Spacetime*, Norton (1976).